\RequirePackage{fix-cm}
\documentclass[smallextended]{svjour3}       
\smartqed  
\usepackage{graphicx}
\usepackage[ruled,vlined,linesnumbered]{algorithm2e}
\usepackage{float}
\usepackage{amsmath}

\usepackage{amssymb}
\usepackage{tcolorbox}
\usepackage{amsfonts}
\usepackage{hyperref}
\usepackage{algorithmic} 
\usepackage{graphicx}
\usepackage{textcomp}
\usepackage{xcolor}
\usepackage{xspace}
\usepackage{multirow}
\usepackage{booktabs}
\usepackage{soul}
\usepackage{balance}
\usepackage{tcolorbox}
\usepackage{array}
\usepackage{multirow}
\usepackage{caption}
\usepackage{subcaption}
\usepackage[inline]{enumitem}
\usepackage[T1]{fontenc}
\usepackage{lipsum}

\begin{document}

\title{When Uncertainty Leads to Unsafety}
\subtitle{Empirical Insights into the Role of Uncertainty in Unmanned~Aerial~Vehicle Safety}


\author{Sajad~Khatiri         \and
        Fatemeh~Mohammadi~Amin \and
        Sebastiano~Panichella \and
        Paolo~Tonella 
}


\institute{ 
            Sajad Khatiri \at Università della Svizzera italiana \& Zurich  University of Applied Sciences, Switzerland\\
    \email{sajad.mazraehkhatiri@usi.ch}           
    \and
            Fatemeh Mohammadi Amin \at University of Zurich \& Zurich  University of Applied Sciences, Switzerland\\
    \email{fatemeh.mohammadiamin@zhaw.ch}  
    \and
            Sebastiano Panichella \at University of Bern, Switzerland\\
    \email{sebastiano.panichella@unibe.ch}
    \and
            Paolo Tonella \at Università della Svizzera italiana, Switzerland\\
    \email{paolo.tonella@usi.ch}  
}

\date{Received: date / Accepted: date}

\maketitle

\begin{abstract}
Despite the recent developments in obstacle avoidance and other safety features, autonomous Unmanned Aerial Vehicles (UAVs) continue to face safety challenges. No previous work investigated the relationship between the behavioral uncertainty of a UAV, characterized in this work by inconsistent or erratic control signal patterns, and the unsafety of its flight. By quantifying  uncertainty, it is possible to  develop a predictor for unsafety, which  acts as a flight supervisor.
We conducted a large-scale empirical investigation of safety violations using PX4-Autopilot, an open-source UAV software platform. Our dataset of over 5,000 simulated flights, created to challenge obstacle avoidance, allowed us to explore the relation between uncertain UAV decisions and safety violations: up to 89\% of unsafe UAV states exhibit significant decision uncertainty, and up to 74\% of uncertain decisions lead to unsafe states.
Based on these findings, we implemented \textsc{Superialist} (Supervising Autonomous Aerial Vehicles), a \textit{runtime uncertainty detector} based on autoencoders, the state-of-the-art technology for  anomaly detection. \textsc{Superialist} achieved high performance in detecting uncertain behaviors with up to 96\% precision and 93\% recall. Despite the observed performance degradation when using the same approach for predicting unsafety (up to 74\% precision and 87\% recall), \textsc{Superialist} enabled  early prediction of unsafe states up to 50 seconds in advance.

\end{abstract}

\section{Introduction}
\label{sec:intro}

Autonomous Cyber-physical systems (CPSs), such as Unmanned Aerial Vehicles (UAVs), are employed in different application scenarios, including crop monitoring \cite{CarboneAMOSKNT21} and surveillance~\cite{DBLP:journals/sensors/BalestrieriDVPT21}, to search and rescue in disaster areas \cite{ChowdhurySMLB21}, which pose significant challenges when ensuring UAV operational safety \cite{DAndrea14,ChowdhurySMLB21,UAV:2022,ZAMPETTI2022111425,ICST-UAV2025}. 
UAV developers enhance the safety measures of the autonomous control modules in various ways, e.g., by incorporating functionalities for autonomous obstacle avoidance, collision prevention, safe takeoff, and landing, or by defining a Geo-fence (i.e., a virtual boundary of the safe flight zone).
Despite such safety features being embedded in both commercial drones and popular open-access UAV Autopilot systems~\cite{ardupilot,px4-Autpilot}, various incidents involving drones \cite{accidents:2019} suggest that UAVs present relevant safety challenges to be addressed by researcher and UAV developers \cite{3711906}. 


Testing UAV functions and features in critical scenarios is still lacking. For this reason, researchers proposed to use simulation environments to test general CPSs and UAVs in a diversified set of scenarios \cite{HuangSLFB21,BojarczukGLDH0S21,PiazzoniCAYSV21,SDCScissor,NguyenHG21,surrealist}, and to support testing automation \cite{AlconTAC21,Wotawa21}, automated test generation \cite{surrealist}, regression testing \cite{Birchler2022Machine,Birchler2022Single} and debugging \cite{SmithR21,RoyHACC21} activities. 
However, while simulation-based testing remains a crucial practice for evaluating the functional and safety requirements before any real-world flight \cite{afzal2018crashing,afzal2021simulation,wang2021exploratory,surrealist}, even with the most advanced testing processes in place, UAVs may deviate from their intended behavior at runtime \cite{surrealist,3571854,ZAMPETTI2022111425,UAV:2022} due to many potentially unforeseen runtime factors such as weather conditions and sensor noise. 
Hence, the current commercial deployment of autonomous systems such as UAVs (legally) requires the presence
of a human pilot/driver to actively monitor their safe operation and be prepared to assume control if necessary~\cite{matternet}.


Autonomous systems (e.g., self-driving cars and UAVs) deal with dynamic, non-stationary, and highly unpredictable operational environments \cite{arnez2020comparison}. 
Since it is practically not feasible to account for all the details of the operational environment,
recent research suggested the notion of \textit{uncertainty} \cite{LaurentKAIV23,stocco2020,mcallister2017concrete,ShinCNSBZ21,SuLHHFDM23,CatakYA22,HanAYAA23,weiss2021fail} to quantitatively measure the degree of doubt or variability in the autonomous systems' perception and decision-making process at runtime. 
This can serve as a way to identify unfamiliar contexts and monitor the reliability of their runtime decisions~\cite{MaAY21,Xu00A22,AsmatKH23} particularly in challenging environments.

So far, Uncertainty quantification has been mostly used in the context of Deep Neural Networks (DNN) that provide a vision component to the system~\cite{mcallister2017concrete,arnez2022towards,stocco2020,weiss2021fail}, considering its contribution to the overall system behavior~\cite{arnez2022towards}, while uncertainty in autonomous systems without DNN components has not been investigated much.
Recently proposed \textit{black-box} uncertainty estimation methods that only consider the sensor input of the autonomous system (e.g., camera feed, sensor signals)~\cite{stocco2020} and/or its final output (e.g., motor control signals, position)~\cite{michelmore2020uncertainty}, are promising and can potentially generalize beyond any specific implementation method (e.g., DNNs).

\begin{figure}
\centering
\includegraphics[width=\columnwidth]{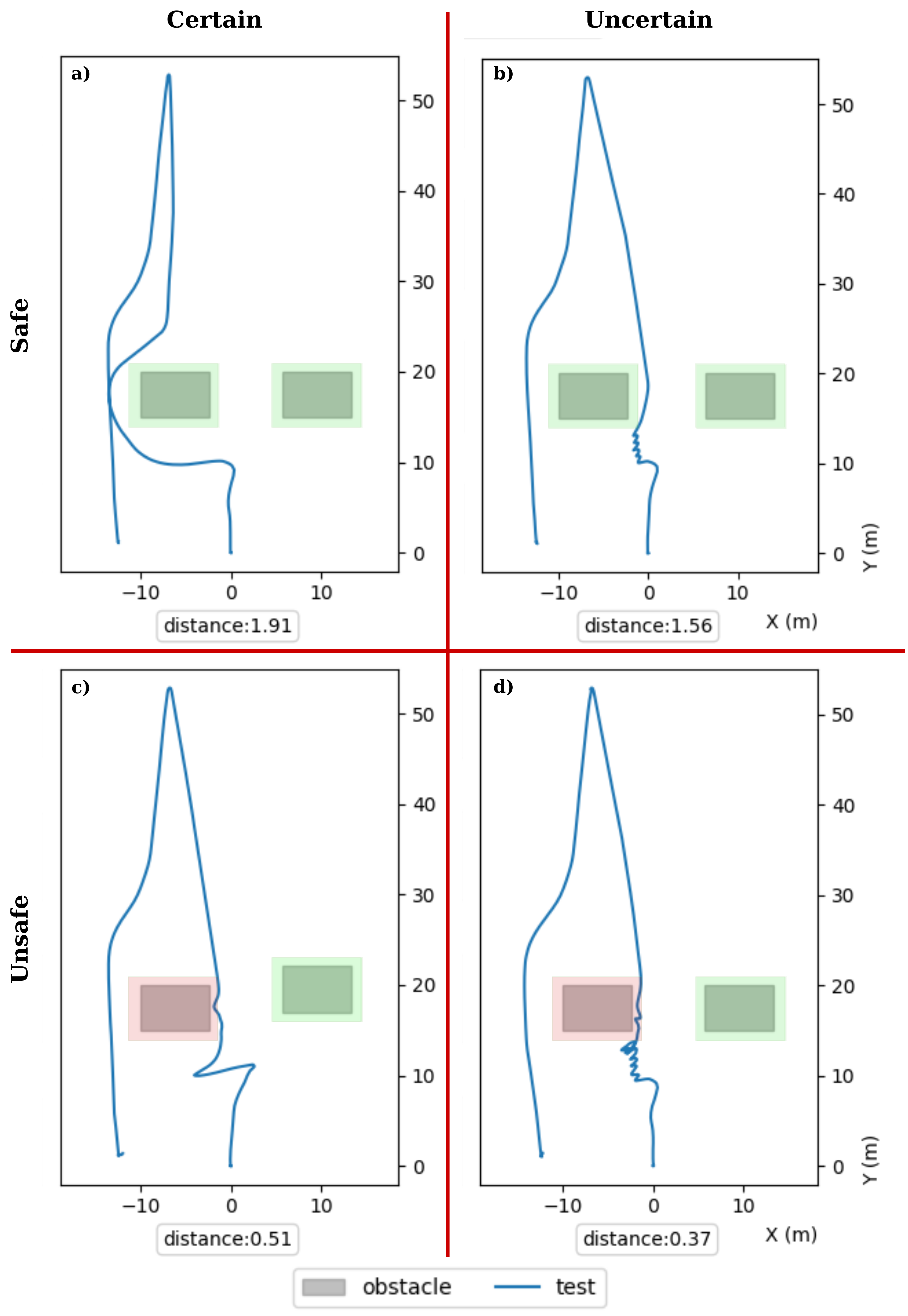}
\caption{Examples of flight safety and certainly levels }
\label{fig:safety-matrix}
\end{figure}

In this context, we focus on the behavioral effects of uncertainty rather than its internal implications and introduce \textit{Decision Uncertainty}, a black-box uncertainty measurement obtained by analyzing the inconsistencies in the control signals (i.e., system output) of the autonomous system, which could potentially lead to unstable, erratic, and unpredictable movements (as shown in Figure \ref{fig:safety-matrix}.b), and, in extreme cases, to safety violations (as shown in Figure \ref{fig:safety-matrix}.d).
While previous studies have proposed approaches for anomaly detection in runtime variables of UAV systems \cite{sindhwani2020unsupervised,TliliACO22,shar2022dronlomaly,titouna2020online}, to the best of our knowledge, no previous study has investigated \textit{decision uncertainty} in the context of autonomous UAV flights, nor proposed a proper quantification method for it. 
Moreover, it remains unclear to what extent the uncertainty quantification aligns with the UAV safety metrics (safe or unsafe UAV states); i.e., whether there are (strong or soft) relations between uncertain UAV decisions and the occurrence of safety violations.

Figure \ref{fig:safety-matrix} shows the flight trajectories of a UAV in 4 similar simulation-based test cases (with marginal differences in the placement of the right obstacle) exhibiting different behaviors. 
Assuming a safe distance of about 1.5m (proposed by previous work~\cite{surrealist}) to the obstacles (green and red highlights around the obstacles), the top simulations (a,b) are considered safe, while the ones at the bottom (c,d) represent unsafe flight trajectories that even lead to crashing into the obstacles (d).
On the other hand, the left-side flight trajectories (a,c) exhibit no or limited signs of uncertainty in UAV's flight path decisions. In contrast, in the ones on the right (b,d), we observe an obvious hesitation and instability in the flight direction and heading angle of the drone right before flying into the risky area in the middle of the two obstacles.  

These observations motivate us to explore the (non-trivial) relationship between uncertainty and unsafety in UAV flights, as well as the extent to which unsafe states can be predicted (and potentially prevented) during the flight.
Hence, our primary goal is to understand and characterize this relation by answering the following general question:
\begin{tcolorbox}
What is the Role of Uncertainty in the Safety States of UAVs? When and Why does UAVs' Uncertainty lead to  Unsafe states? 
\end{tcolorbox}
To answer this question, we conducted the first large-scale data-driven empirical investigation of UAV safety violations and their relation to decision uncertainty.
We considered PX4-Avoidance~\cite{px4-avoidance}, an open-source and widely used autonomous obstacle avoidance system built on top of PX4-Autopilot~\cite{px4-Autpilot}, one of the most popular open-source UAV Autopilot systems and firmware, as a relevant use case. 
Employing evolutionary algorithms~\cite{surrealist}, we simulated UAV flights in diverse environments and flight missions, revealing potential misbehaviors, crashes, and safety violations. Our extensive dataset features about 5,000 simulated UAV flights, including unsafe and uncertain conditions. 
Manual analysis of a subset enabled exploration of the link between uncertain UAV decisions and safety violations.

In our simulated flights, we discovered a moderate correlation between UAV safety and the level of \textit{black box uncertainty} (i.e., uncertainty measured only by observing the high-level UAV control signals, without considering the internal operation of the obstacle avoidance module). 
Approximately 64-89\% (calculated over two distinct datasets) of unsafe UAV states displayed substantial decision uncertainty prior to the incident, indicating that 11-36\% of unsafe states did not show significant signs of uncertainty.
On the other hand, only 50-74\% of uncertain decisions are eventually leading to unsafe states, undermining their ability for unsafety prediction.

To explore the potential of predicting safety violations with minimal UAV black box data, we employed a \textit{runtime uncertainty detector using autoencoders}. This system, named \textsc{Superialist} (Supervising Autonomous Aerial Vehicles), utilizes a minimal set of external, black box control signals to model the normal UAV behavior and identify uncertain behavior at runtime. Our experiments revealed that \textsc{Superialist} is highly effective at detecting uncertain UAV decisions achieving a precision of 90-95\% and a recall of 84-93\%.
Moreover, on average, the first uncertain state was detected about 46 seconds before the drone reached a critical zone (less than 1m distance to the obstacles), when the drone was on average about 3.6 meters away from the obstacles, opening a very large time and distance window for reaction.
However, precision and recall decline notably when using the same approach for unsafety prediction. This aligns with the correlation between unsafety and black box uncertainty observed in our empirical study. 

Our experimental results provide two important findings: (1) \textit{uncertain decisions} may eventually lead to safety violations, but there are notable exceptions; (2) black box metrics are very accurate predictors of \textit{uncertain behaviors}; (3) however, a non-negligible proportion of unsafe states are not associated with signs of uncertainty, pointing to the need for future work on misbehavior prediction based on combinations of black box and white box indicators.

Our contributions complement previous research as follows:

\begin{itemize}
    \item  We conduct the first large-scale data-driven empirical investigation of UAV safety violations, studying the relation between uncertain and unsafe states of UAVs. 
    \item We propose and evaluate \textsc{Superialist}, a \textit{black-box} approach for detecting UAV \textit{decision uncertainty} at runtime, utilizing a small set of UAV control signals.
    \item Based on the identified correlation of uncertain UAV decisions and upcoming unsafe states, we evaluate \textsc{Superialist} as an unsafety prediction approach.
    \item We discuss relevant insights of our work and publicly share the generated datasets and the code to reproduce our results in the replication package~\cite{RP}. 
\end{itemize}

\section{Background}
\label{sec:background}
\subsection{PX4 Platform}

PX4~\cite{px4-Autpilot} is a comprehensive open-source software platform for developing UAV systems, including autopilot software, obstacle avoidance algorithms, ground control station tools, multiple simulation environments, flight log analysis tools, and support for various hardware and mechanical setups. 
PX4 is used in a wide range of use cases, from consumer drones to industrial applications, and is the leading research platform for drones. 
This section explores relevant PX4 features to our study.

\subsubsection{Simulation Environments}
Simulators, such as Gazebo \cite{gazebo}, enable PX4 to control a virtual vehicle and receive sensor data from the simulated environment. The UAV pilot interacts with the virtual vehicle (or the physical vehicle) through tools such as a ground control station (GCS), a radio controller (RC), or an offboard API (e.g., ROS). 
These interfaces send control commands and receive telemetry data. PX4 is compatible with various Hardware-in-the-Loop (HIL) and Software-in-the-Loop (SIL) simulators \cite{PX4-simulators}, enabling safe testing of control algorithms before real flights, to reduce the risk of real vehicle damage. 
In the current study, we use the Gazebo simulator \cite{gazebo} to assess PX4's obstacle avoidance feature.

\subsubsection{Autonomous Flight \& Obstacle Avoidance}
    \label{sec:backgd-flightmodes}

\begin{figure}
\centering
\includegraphics[width=0.8\columnwidth]{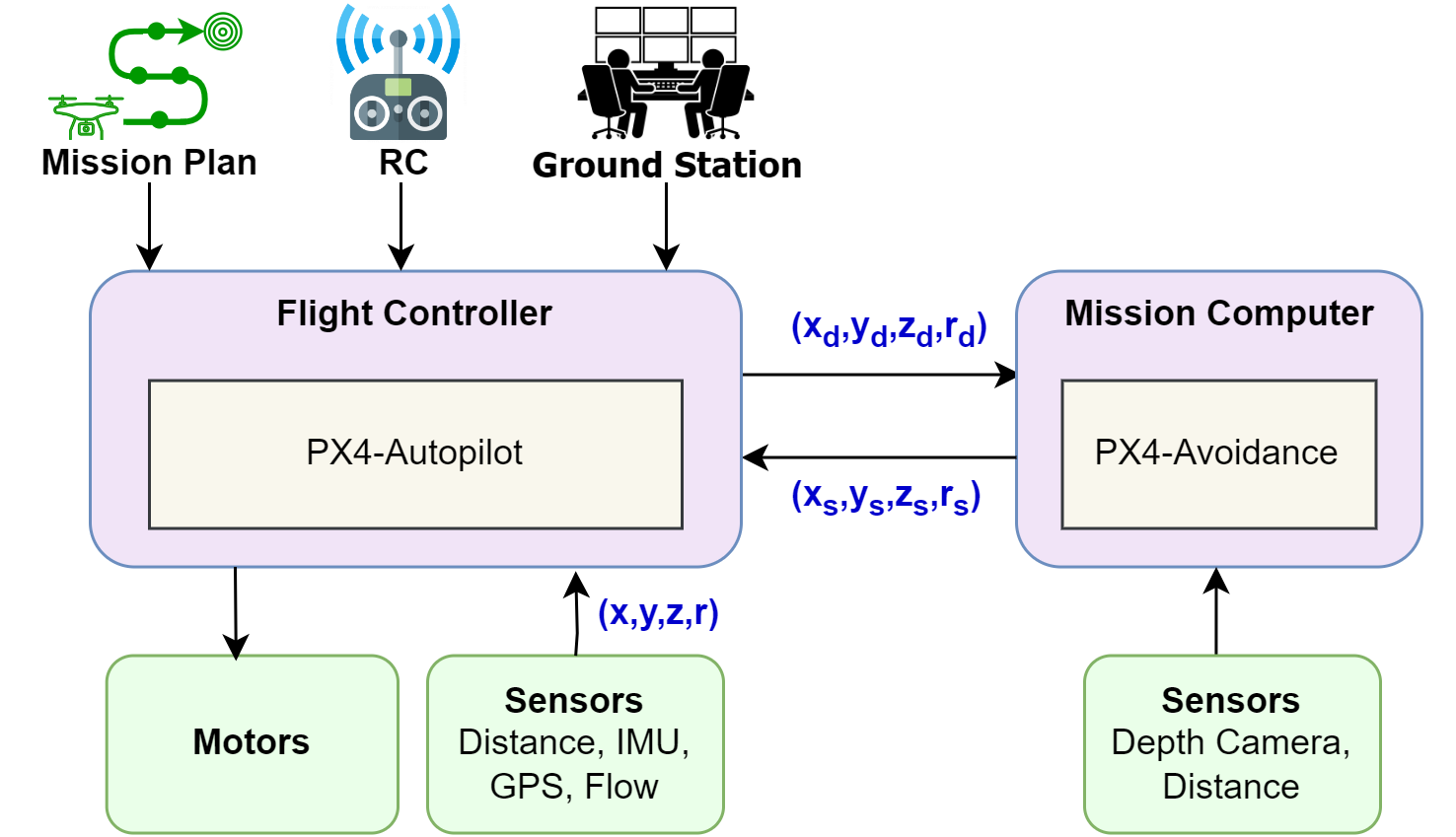}
\caption{PX4 Platform Architecture }
\label{fig:px4}
 \vspace{-2mm}
\end{figure}

PX4 supports various \textit{flight modes} offering different autonomy levels. The \textit{mission mode} enables fully autonomous flight, executing predefined tasks with minimal human intervention. A \textit{mission plan} is a structured set of flight waypoints and instructions for UAVs, applicable in tasks like aerial surveys and mapping operations.
As demonstrated in Figure \ref{fig:px4}, PX4-Autopilot oversees drone autonomy and collaborates with PX4-Avoidance — a separate module featuring intelligent camera-based obstacle detection and avoidance. PX4-Avoidance assesses real-time surroundings during mission execution using onboard cameras and distance sensors, ensuring obstacle-free navigation~\cite{baumann2018obstacle}. 
The system continuously receives planned (desired) trajectory waypoints, denoted as $\langle x_d, y_d, z_d, r_d\rangle$, representing spatial coordinates ($x, y, z$), and heading orientation ($r$) from PX4-Autopilot. It checks if these points would bring the drone too close to obstacles. If a potential collision is detected, PX4-Avoidance updates the waypoints to $\langle x_s,y_s,z_s,r_s\rangle$, allowing the drone to find a safe path through obstacles iteratively. This functionality is crucial for UAVs to autonomously navigate complex environments, ensuring their safety and precise mission execution.

\subsubsection{Flight Logs}
    \label{sec:backgd-logs}

PX4 logs all communication between the RC/GCS and UAVs, as well as internal communication between UAV modules, including sensor outputs, GPS data, runtime commands, motor control signals, mission points, desired waypoints by PX4-Autopilot, and updated waypoints by PX4-Avoidance \cite{px4-log-analysis}. These logs aid post-flight analysis, troubleshooting, and real-time monitoring of UAV behavior (an example of flight log at the  link\footnote{\href{ https://logs.px4.io/plot_app?log=f986a896-c189-4bfa-a11a-1d80fa4b9633}{https://logs.px4.io/plot\_app?log=f986a896-c189-4bfa-a11a-1d80fa4b9633}}).

There are several factors influencing a drone's path in autonomous flight. In this study, we focus on monitoring obstacle avoidance in a \textit{black-box} manner, primarily examining the inputs and outputs of the PX4-Avoidance module, as illustrated in Figure \ref{fig:px4} (with $ts$ denoting a timestamp):
\begin{itemize}
    \item \textit{Desired (Blind)} Waypoints (<ts, x$_d$, y$_d$, z$_d$, r$_d$>): PX4-Autopilot calculates optimal waypoints at each timestep to reach future mission points, disregarding obstacles.
    \item \textit{Safe (Control)} Waypoints (<ts, x$_s$, y$_s$, z$_s$, r$_s$>): PX4-Avoidance updates the \textit{desired waypoint} using camera-obtained obstacle information to plan an obstacle-free path. PX4-Autopilot adjusts the drone's movement accordingly.
    \item \textit{Local Position (<ts, x, y, z, r>):} Actual drone position data (i.e., flight trajectory) relative to takeoff, computed from GPS and other sensors. 
\end{itemize}

The heading (r) denotes the drone head rotation angle from the startup state. This is crucial as our drone, similar to many commercial ones, relies on a single front-facing stereo camera for obstacle detection. Rotating (i.e., changing the heading angle) is essential for obtaining a comprehensive view of the environment in challenging positions near obstacles.

\subsection{Flight Quality Attributes}

\subsubsection{Decision Uncertainty}
\label{sec:background-uncertainty}

Ensuring the safety of autonomous systems requires the identification of unfamiliar contexts by modeling their uncertainty~\cite{mcallister2017concrete}. We can generally distinguish between two sources of uncertainty that affect the system's performance:
\textit{Aleatoric uncertainty} accounts for the uncertainty truly present in the nature of the operational domain and can be seen as the randomness and ambiguity in the surrounding environment or the incapability of completely sensing all the details of the environment~\cite{kendall2017uncertainties}. 
\textit{Epistemic Uncertainty}, also known as \textit{model uncertainty}, accounts for uncertainty in the ML model parameters, or more generally, in the control algorithm, which descends from lack of knowledge (e.g., inadequate training set). 
This type of uncertainty captures  the system's lack of knowledge about certain operational scenarios due to insufficient training (or implementation) 
\cite{kendall2017uncertainties}. 
In safety engineering ---the application of engineering principles and methods to identify, analyze, and control hazards within a system throughout its lifecycle to achieve an acceptable level of safety and prevent harm--- uncertainty often refers to the lack of knowledge about factors contributing to risk (i.e., epistemic uncertainty), such as the precise failure rate of components, the exact environmental conditions, or the completeness of system models~\cite{ryan2024dynamic,denney2015dynamic,hawkins2013assurance}. 
%
%

While we acknowledge this established view, our study focuses on a different, albeit related, aspect: the observable behaviors arising when an autonomous system encounters situations challenging its control logic, potentially stemming from underlying epistemic or aleatoric uncertainty. We name this phenomenon \textit{Behavioral Control Uncertainty} (or simply \textit{Decision Uncertainty} within the context of our study --  to clarify the differences and commonalities between this definition and the existing ones) and define it operationally based on the system's output:

\begin{tcolorbox}[colback=gray!15!white,colframe=black, boxsep=2pt,left=3pt,right=3pt,top=3pt,bottom=3pt]
\textbf{Decision Uncertainty}: \textit{``The likelihood that the autonomous control system takes inconsistent, unsteady, or even contradicting control decisions (i.e., movement instructions) in a short period of time or specific section of the mission. 
Symptoms may include deviation from the optimal path, instability, abnormal and hazardous frequent changes in the movement pattern, and even crashing into nearby obstacles''.}


\end{tcolorbox}

\begin{figure}
\centering
\includegraphics[width=0.75\columnwidth]{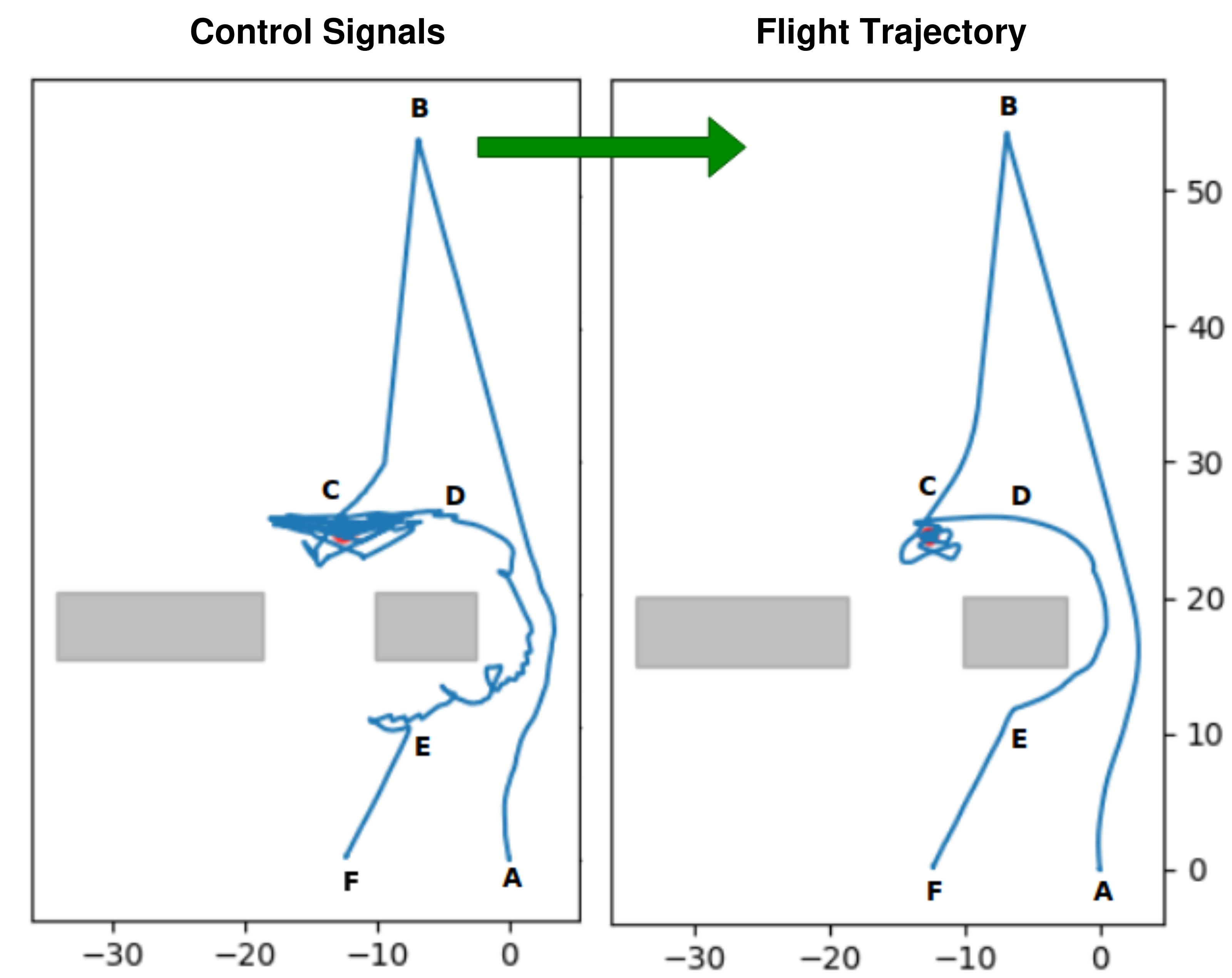}
\caption{The control signals (waypoints) that the drone is instructed to follow and the resulting flight trajectory}
\label{fig:waypoints}
\end{figure}


Intuitively, we characterize this behavioral uncertainty by analyzing the stability and consistency of the drone's planned path (control signals), rather than directly measuring uncertainty about internal parameters or external risk factors. Our rationale for this black-box, output-focused approach is its potential for real-time monitoring, using readily available control data, and its ability to capture system issues in complex scenarios, regardless of the specific internal cause. This study empirically investigates the extent to which this observable behavioral uncertainty correlates with, and potentially predicts, unsafe flight states.

Figure \ref{fig:waypoints} demonstrates the relation between the high-level UAV control signals (i.e., safe waypoints that the drone is instructed to follow; left picture) and the resulting flight trajectory (i.e., actual positions of the drone derived from sensors; right picture) during an autonomous mission ($ A \rightarrow B \rightarrow F$) with the presence of two obstacles. While the control waypoints in the initial sections of the flight ($A \rightarrow B \rightarrow C$) are consistent and the resulting flight trajectory is closely following them, they get inconsistent and deviate from the optimal path as the drone enters the critical area around the obstacles ($C \rightarrow D \rightarrow E$). The resulting flight trajectory also diverges from the optimal path. While in this case the decision uncertainty (especially around $C$) is observable even in the resulting flight trajectory, it is generally much more evident and easily measured in the UAV control signals (i.e., safe waypoints).


\subsubsection{Flight Safety}
\label{sec:background-safety}
UAV safety encompasses various requirements. This study specifically emphasizes the obstacle avoidance system's reliability in autonomous flight.
This includes the following requirements: 
The drone must avoid crashing into obstacles (\textbf{Requirement 1}) and maintain a \textit{safe distance} from surrounding obstacles to prevent potential crashes in the event of sudden environmental changes (e.g., wind) (\textbf{Requirement 2}).
Previous study~\cite{surrealist} suggested \textit{1.5m} distance as the safety threshold (highlighted areas around the obstacles in Figure \ref{fig:safety-matrix}) considering the size of the case study drone. To enable a more precise safety assessment, we manually label the safety aspects of the flights as described in Section \ref{sec:labeling} instead of using a fixed threshold. Then, we propose a runtime unsafety prediction method in Section \ref{sec:methodology-rq2}.


\subsubsection{Behavioral Non-determinism}
\label{sec:background-nondeterminism}

Many of the components in the UAV hardware and sensors, software platform, obstacle avoidance, and path planning algorithm, as well as simulation environment, have some degrees of randomness involved. 
Hence, non-determinism is always expected in a flight scenario \cite{surrealist,aerialist}.
If a test case is simulated multiple times, sometimes slight and sometimes considerable differences in the flight trajectories, as well as in the decision certainty and the flight safety, can be observed. 
This is prevalent in corner cases, where a test scenario passes in some runs, but fails in others (these are also called \textit{flaky tests}~\cite{LuoHEM14,AminiNN24}), or the drone takes different paths around the obstacles in different runs. 
Such corner cases are also the most interesting for our empirical investigation, as we want to understand whether uncertainty and unsafety correlate in these cases. Section \ref{sec:dataset} describes how we deal with non-determinism in our dataset generation process, in order to obtain statistically valid results despite the presence of non-deterministic behaviors.


\section{Related Works}
\label{sec:related}
\subsection{Safety Assurance for UAVs}

The traditional aviation industry adheres to well-established and rigorous safety standards governing system development and safety assessment processes \cite{SAE:ARP4754A,SAE:ARP4761}. These standards mandate comprehensive design assurance and verification activities scaled according to potential risks, forming the basis for demonstrating aircraft safety. 
In contrast, the rapidly evolving field of consumer and small UAVs currently operates under less comprehensive and standardized design assurance frameworks. 
While regulations ---such as the EASA drone classification system (C0-C4) \cite{EASA:Reg2019_945}, drone operational categories (Open, Specific, Certified) \cite{EASA:Reg2019_947}, and CE marking requirements \cite{EU:RED2014_53})--- exist concerning specific technical capabilities, operational limitations, and general product safety for consumer drones \cite{EUDronePort:C1Reqs}, they typically do not enforce the same depth of system development and safety assessment rigor found in traditional aviation. 
This difference highlights a potential gap and an ongoing challenge in establishing and demonstrating sufficient safety assurance for autonomous UAVs, particularly as their operational complexity increases and they interact more with dynamic environments.

%

\textit{Safety assurance} is the process of providing justified confidence that a system meets its defined safety requirements and is acceptably safe for its intended purpose within a specific operational context \cite{hawkins2013assurance}. Traditionally, this process leads to the creation of a safety case: a structured argument, supported by a body of evidence (e.g., analysis results, test reports, simulations), demonstrating that the system's risks are tolerable \cite{hawkins2013assurance}. Safety cases typically reflect the system understanding and safety assessment performed at design time. However, for autonomous systems such as UAVs operating in dynamic and unpredictable environments, this static design-time assurance may become insufficient or invalidated during operation as underlying assumptions change~\cite{ryan2024dynamic,denney2015dynamic}.

To address this challenge, the concept of Dynamic Safety Assurance (DSA) has emerged, advocating for continuous, through-life safety management ~\cite{ryan2024dynamic,denney2015dynamic}. DSA frameworks emphasize the need to monitor the system and its operational context at runtime, continuously assess risks and the validity of safety arguments, and potentially adapt the system behavior to maintain safety. Mechanisms like Dynamic Safety Cases (DSCs) have been proposed to structure this process, integrating runtime evidence to dynamically update the estimated confidence levels in safety claims \cite{ryan2024dynamic}. A cornerstone of DSA is the development of effective runtime monitors capable of detecting anomalies, performance degradation, or emerging hazards \cite{ryan2024dynamic}. Our work contributes to this area by investigating decision uncertainty as a runtime indicator of potential safety issues and proposing SUPERIALIST, a black-box monitor designed to detect such uncertainty, thereby providing evidence that could be leveraged within a DSA framework.

\subsection{Uncertainty Quantification in Autonomous Systems}
Recent studies focused on addressing uncertainty in the context of autonomous systems \cite{LaurentKAIV23,stocco2020,ShinCNSBZ21}, with strategies assessing (or predicting) uncertainty for CPSs~\cite{ShinCNSBZ21,SuLHHFDM23,CatakYA22,HanAYAA23,WeissT21a}, or proposing self-healing approaches for them~\cite{MaAY21,Xu00A22,AsmatKH23}. 
So far, uncertainty (or confidence) quantification has been addressed mostly in the context of DNNs used in perception tasks (e.g., vision components) of autonomous systems.
Bayesian Neural Networks (BNN) provide a rigorous framework for modeling and capturing uncertainty ~\cite{mcallister2017concrete}. 
Although BNNs can be used to implement the perception components of UAVs, the downstream UAV components of the navigation pipeline (i.e., planning and control) typically either ignore the uncertainty from earlier components or do not take uncertainty into account in their predictions and the overall system output does not reflect the internal uncertainty~\cite{arnez2022towards} in any way.
Some recent works have proposed solutions to propagate the internal uncertainty to the output in specific BNN architectures~\cite{arnez2022towards}. However, the general problem of capturing the uncertainty in the system output remains open.

Michelmore \emph{et al.}~\cite{michelmore2020uncertainty} introduce the notion of \textit{Probabilistic Safety}: the probability that a BNN controller will keep the car safe in a certain driving scenario, estimated over multiple simulations with different environmental conditions.
Their safety metric, focused on the behavioral side-effects of uncertainty rather than the internal implications, opens the discussion for \textit{black-box} uncertainty estimation methods that could potentially generalize beyond a specific control method.
Black-box uncertainty can be estimated by considering the sensor input of the autonomous system (e.g., camera feed, sensor signals) and/or its final output (e.g., motor control signals, position), completely ignoring the implementation details.
Stocco \emph{et al.}~\cite{stocco2020} propose an approach for estimating the confidence of DNNs in response to unexpected execution contexts, i.e., camera input. Their goal is to predict potential safety-critical misbehaviors such as out-of-bound episodes or collisions for vision-based autonomous driving systems. 
They utilize autoencoders to build a reconstruction model for the training images of the DNN (i.e., familiar context), and identify unexpected context by monitoring the reconstruction loss of the real-time camera images.

We complement previous studies by introducing \textit{Decision Uncertainty}, A black-box uncertainty measurement due to the inconsistencies in the control signals (i.e., system output) of the autonomous system, potentially leading to unstable, erratic and unpredictable trajectories, and in extreme cases, to safety violations.
Moreover, we complement previous experimental works on uncertainty in CPSs by conducting the first empirical study on the alignment between uncertainty quantification and unsafe UAV states during flights.

\subsection{Anomaly Detection in UAV Components}
Researchers in the fields of robotics and UAVs have previously explored methods to identify anomalies in the behavior and performance of individual drone components such as sensors \cite{9653036}, motors \cite{lu2017motor}, network and communications \cite{YangHYW23,HaoYY23,ahn2020deep}, and their mechanical stability \cite{bektash2020vibration}. 
Galvan \emph{et al.}~\cite{9653036} introduced a CNN (Convolutional Neural Network) based classification system to distinguish normal and anomalous real-time IMU (Inertial Measurement Unit) sensor data. They integrate a cybersecurity plugin into the simulation environment to inject anomalies in the sensor signals and generate a labeled dataset for training and evaluation of the model. 

Other studies regarding anomalies at system-level, focused on identifying the abnormal patterns in inputs (e.g., sensor readings) and the outputs (e.g., drone flying behavior) using intrusion or anomaly detection approaches  \cite{sindhwani2020unsupervised,TliliACO22,shar2022dronlomaly,titouna2020online}. 
Sindhwani \emph{et al.}~\cite{sindhwani2020unsupervised} present an anomaly detection framework for autonomous delivery missions. The framework, trained on sensor and control signals from normal mission executions, identifies artificially injected anomalies representing disabled actuators, invalid control commands, or high wind speed in the flight logs (offline). 
Shar \emph{et al.}~\cite{shar2022dronlomaly} propose an LSTM (Long Short-Term Memory) based approach for detecting anomalies in the sequence of flight states that potentially lead to the drone's physical instability. Their model, trained on sensor data sequences of normal flight logs, is able to detect artificially injected faulty values, reflecting faulty sensor readings, actuator outputs, GPS and Magnetic field interferences, and communication signal losses.
Some studies have also investigated approaches for monitoring drones in swarm applications \cite{ahn2020deep} consisting of multiple collaborating drones. 

Our work complements previous studies by doing the first empirical study on the relationship between the uncertainty in UAV decisions and the safety aspects of the flight. 
Our anomaly detection method proposed for uncertainty detection differs from the above studies by being completely back-box and not using any internal variables (sensor reading, motor signals, GPS), relying only on high-level control commands from an external path planing module. This makes our approach ideal for monitoring high-level path planning algorithms (e.g., obstacle avoidance) external to the main auto-pilot system.   
Additionally, our method is evaluated using faults observed by the obstacle avoidance module in specific obstacle allocations, not synthetic faults injected into sensor values.

\subsection{Simulation-based Testing of Autonomous Systems}
Previous studies have characterized the software bugs in autonomous systems including UAV Autopilot platforms \cite{GarciaF0AXC20,UAV:2022,ZAMPETTI2022111425} and investigated the challenges associated with testing these systems in realistic environments \cite{garcia2020simulation,afzal2020study,surrealist,WuXZL23,MaAY21,VR-study}. 
Afzal \emph{et al.}~\cite{afzal2020study} focused on testing challenges in robotic systems, emphasizing the complexity of test environments.  
Complementary to this, Lidval \emph{et al.}~\cite{lindvall2017metamorphic} and Afzal \emph{et al.}~\cite{afzal2021mithra} developed automated testing frameworks for drones in simulation to address the test oracle problem. 
Woodlief \emph{et al.} introduced PHYS-FUZZ~\cite{elbaum2021fuzzing}, a fuzzing approach for mobile robots that considers their physical attributes and hazards. 
Hildebrandt \emph{et al.}~\cite{elbaum2021world} proposed a "world-in-the-loop" mixed-reality approach for UAV testing that integrates sensor data from both simulated and real-world sources.
In our previous work~\cite{surrealist}, we addressed the problem of realism in simulation-based testing for UAVs by proposing Surrealist, an approach that faithfully replicates real-world test scenarios in simulation environments and generates more challenging test case variants. 
Complementary, we developed a UAV test bench and test generation platform called Aerialist ~\cite{aerialist}, and on top of it we fostered research in this domain by organizing the first UAV test generation competition at the Intl. Workshop on Search-Based and Fuzz Testing (SBFT)~\cite{uav-competition,ICST-UAV2025,SBFT-UAV2025}.

On the other hand, in the existing DL testing literature \cite{ZohdinasabRGT23}, the majority of related works focus on automated test data generation, with only a limited number of input generators being model-based and revealing failures within a simulated environment. 
Recent studies~\cite{stocco2020,SDCScissor,Birchler2022Machine,Birchler2022Cost2,Birchler2023Teaser,AbdessalemNBS18} proposed approaches to test self-driving cars by creating scenarios that induce failures in the driving process.

While such testing practices are aimed at ensuring the reliability of the system under test in diverse sets of scenarios and environments, the safety criticality of the UAV applications enforces the additional need to monitor them at runtime to ensure their proper behavior\cite{matternet}.
Similar to such prior studies, we employ simulators to assess the behavior of autonomous systems and complement them by (i) empirically exploring the relationship between uncertainty and safety states of UAVs. Additionally,  (ii) we introduce a black-box runtime uncertainty detection tool using autoencoders.

\section{Methodology}
\label{sec:methodology}
\label{sec:approach}

This section discusses the goal of our study, our research questions, and the research methods behind each question.
Specifically, we focus on three main research questions (RQs), which are individually detailed in the following sections:  

\paragraph{\textbf{RQ$_1$ [Unsafety vs. Uncertainty]}} 
\textit{Do Unsafety and Uncertainty States of UAV flights correlate?}

\noindent We aim to investigate the degree of correlation between the safety (or unsafety) of UAV flights and the uncertainty in their decision-making. 
This is crucial in evaluating how uncertainty in flight behavior can be used for unsafety prediction.

\paragraph{\textbf{RQ$_2$ [Uncertainty Detection]}} 
\textit{To what extent is it possible to detect UAVs' Uncertainty during the flight?}

\noindent We aim to assess the feasibility and accuracy of real-time uncertainty detection for UAVs during the flight using anomaly detection technologies.

\paragraph{\textbf{RQ$_{3}$[Unsafety Prediction]}} 
\textit{To what extent can we predict unsafe UAV states during the flight using uncertainty detection?} 

\noindent Building on RQ$_1$ and RQ$_2$, we focus on the challenge of predicting unsafe UAV states during flight based on uncertainty detection. The aim is to evaluate if and how early uncertainty identification can serve as a mechanism for preventing unsafe conditions.

\subsection{RQ$_1$ - Unsafety vs. Uncertainty}
This research question is designed to increase our understanding of safety and certainty (aka \textit{confidence}) in the context of autonomous UAV flights. 
This section describes our method for generating diverse simulated flight scenarios and outlines the guidelines and steps for manually analyzing and labeling them regarding flight safety and decision certainty. It also discusses the metrics measuring the correlation between uncertainty and unsafety.

\subsubsection{Flight Datasets}
\label{sec:dataset}

An essential prerequisite for our analysis is acquiring a comprehensive dataset of flight logs covering various scenarios, including the four flight conditions shown in Figure~\ref{fig:safety-matrix}: \textit{safe}, \textit{unsafe}, \textit{certain}, and \textit{uncertain} behaviors. 

To generate the required datasets, we used \textit{Surrealist}\footnote{https://github.com/skhatiri/Surrealist}, an automated simu\-lation-based test case generation tool for UAVs that we developed in previous work~\cite{surrealist}. \textit{Surrealist} is compatible with autonomous PX4 missions and generates challenging environments with obstacles, making it difficult for PX4-Avoidance to identify a safe path. Using a search-based approach, the tool iteratively introduces static box-shaped obstacles into the simulation environment, adjusting their size, position, and orientation. \textit{Surrealist} also enables multiple simulations of each test case, allowing for the analysis of potential non-deterministic behaviors and minimizing their impact during the test generation process.
Figure \ref{fig:safety-matrix} illustrates test cases generated by \textit{Surrealist}, showing iterations where small adjustments to the placement of the obstacle on the right lead the drone onto an unsafe path. For our study, we included a dataset generated by the original \textit{Surrealist} implementation (test2\_ds, as explained later) among our datasets.

While the initial \textit{Surrealist} setup effectively generates \textit{realistic} failed test cases for obstacle avoidance~\cite{surrealist}, these cases lack diversity and focus primarily on safety issues without explicitly targeting uncertainty. To address this, we extended the \textit{Surrealist} framework to generate new flight datasets featuring more diverse scenarios, enabling a more robust correlation analysis between safety and certainty. We focus exclusively on scenarios with static obstacles, as the system under test (PX4-Avoidance) is not designed to handle complex dynamic situations involving moving obstacles. These datasets can also be used as diverse training data for anomaly detection approaches. 

\paragraph{Distance Measure.}

\textit{Surrealist}'s search process aims to minimize a predefined distance measure (e.g., the minimum distance between the drone and obstacles) over multiple iterations. The original implementation promotes scenarios (i.e., obstacle placements) where the UAV flies in the middle of two obstacles, and makes it iteratively more challenging by closing the gap between the obstacles. Our improved distance metric not only promotes the UAV's proximity to obstacles but also prioritizes test cases with non-deterministic flight trajectories. Non-deterministic trajectories involve varying paths in multiple runs, assuming increased uncertainty in the autopilot system. Our experiments confirmed that non-deterministic test cases exhibit more signs of uncertainty.
Formally, we introduce the new distance function, which is minimized by \textit{Surrealist}'s search process, as follows:

    \begin{align}
    \begin{split}
    & \text{\textit{sum\_dist}} = \min_{p \in \text{trj.points}} \sum_{o \in \text{obs}} d(p, o) \\
    & \text{\textit{ave\_dtw}} = \frac{1}{n}\sum_{t \in \text{trjs}} \text{\textit{dtw(t, ave\_trj)}}\\
    & \text{\textit{distance}} = 
    \begin{cases}
    \text{\text{\textit{sum\_dist}} - \textit{ave\_dtw}} & \text{if}~ \textit{ave\_dtw} > \textit{MAX\_DTW} \\
    \text{\text{\textit{sum\_dist}}} & \text{otherwise}
    \end{cases}
    \end{split}
    \end{align}
    
    \emph{sum\_dist} (as in the original implementation~\cite{surrealist}) accounts for the minimum distance of a single trajectory point to all of the obstacles combined (favoring the generation of scenarios with obstacles that are close to each other). The  term ($\sum_{o \in \text{obs}} d(p, o) $) represents the sum of the distances of a trajectory point ($p$) to all of the obstacles($o$) present in the environment.
    \emph{ave\_dtw} accounts for the average DTW distance \cite{berndt1994using} between the average flight trajectory (\emph{ave\_trj}, among multiple test executions) and each of the individual flight trajectories.
    The latter is a new term, accounting for the non-determinism in the flight path of multiple simulations of the test case. 
The primary objective is to maximize non-determinism while minimizing obstacle distance, accomplished by optimizing the final \emph{distance} formula.
 Since it is always possible to observe some trajectory differences~\cite{surrealist}, 
 we decided to consider the \emph{ave\_dtw} term only if it is bigger than a predefined \emph{MAX\_DTW} value (65 in our experiments; determined empirically).

\begin{figure}
\centering
\includegraphics[width=\columnwidth]{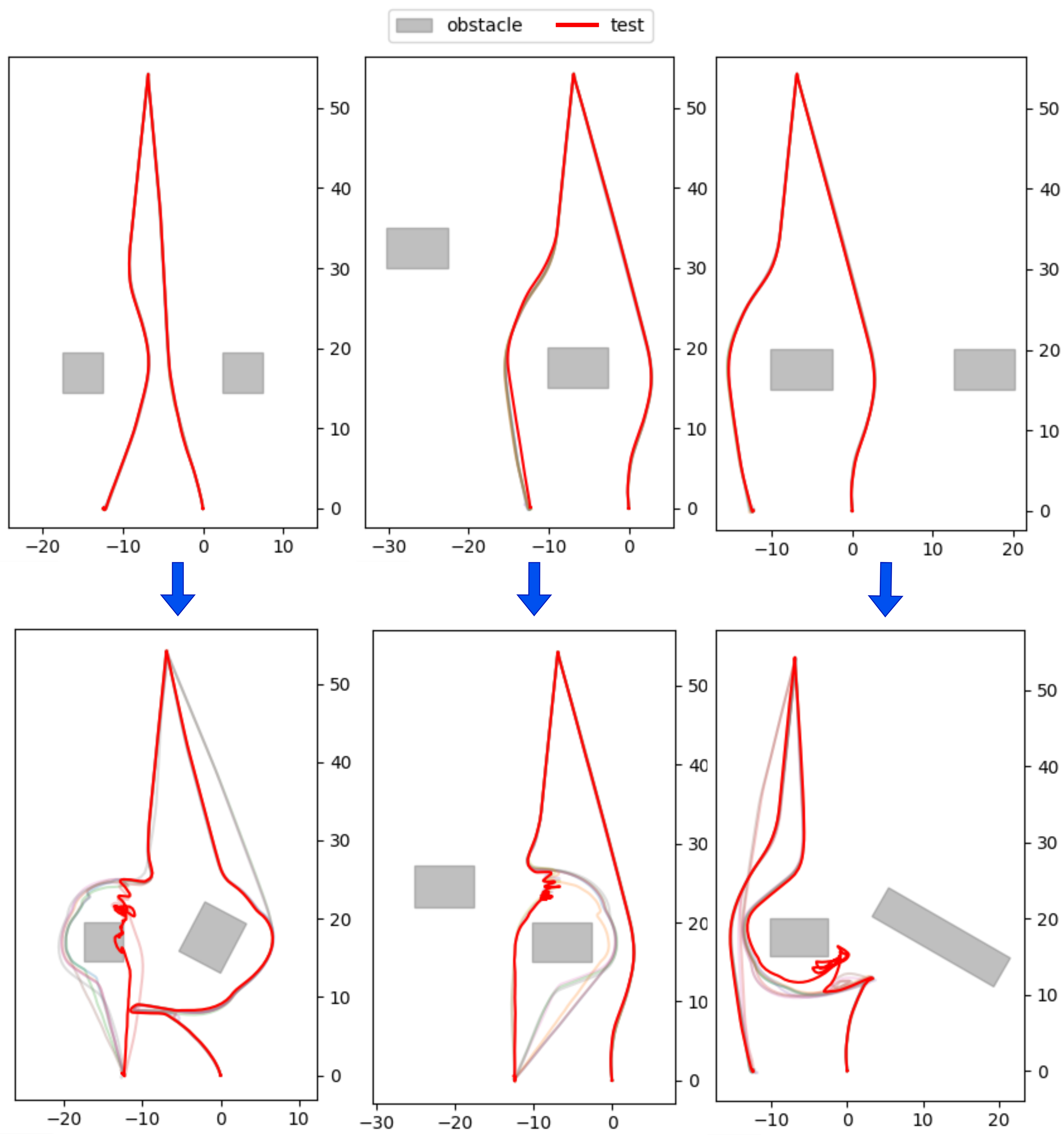}
\caption{Seed test cases (top) and their sample generated tests (bottom)}
\label{fig:seeds}
\end{figure}
\paragraph{Search Process.}
In the original \textit{Surrealist}~\cite{surrealist} setup, obstacles are limited to movement in the $x,y$ plane for size realism. Our approach expands this by enabling the search algorithm to adjust obstacle size (length and width) and rotation, enhancing test diversity. Increased randomization in mutation operator selection at each iteration introduces further diversification of test scenarios.
To further enhance diversity, we generate test cases for 6 distinct initial obstacle positions in a predefined mission plan, concurrently executing each test case 9 times, resulting in approximately 5,000 flight logs. 
Figure \ref{fig:seeds} illustrates 3 test generation seeds (top) and a sample generated test case using them as seed in the search process. During each search, one of the obstacles is fixed, and the other one is iteratively resized, rotated, and moved to minimize the defined distance metric. Non-deterministic behaviors are apparent from the flight path differences among the 9 executions in the bottom plots. 
The 6 resulting datasets were compared, and the most diversified one, consisting of 107 different test cases and 956 flight logs, was selected for further analysis (see Section \ref{sec:preprocess}).


\subsubsection{Manual Flight Labeling}
\label{sec:labeling}
To investigate the degree of correlation between unsafety of UAV flights and the uncertainty in their decision-making, and to assess \textsc{Superialist}'s ability in uncertainty detection and unsafety prediction, a labeled dataset with reliable ground truth labels is required. Below, we outline the manual labeling process employed for UAV flight scenarios.

\paragraph{Labeling Tool.}
We created a tool with a visualization similar to Figure \ref{fig:safety-matrix} to aid flight labeling. It analyzes flight logs, generates simulated trajectory plots with obstacle indicators, and emphasizes the drone's minimum distance to obstacles. Users can conveniently review and label flights as 'safe' or 'unsafe,' and evaluate the UAV's decision certainty by marking it as 'certain' or 'uncertain' following the below guidelines.

\paragraph{Safety Assessment.}
    Following the safety requirements mentioned in Section \ref{sec:background-safety}, 
    maintaining a fixed minimum distance to the obstacles (e.g., 1.5m safety threshold~\cite{surrealist}) during the flight could be the most important criterion for safety assessment. However, during the labeling process we realized that the safety assessment of a drone flight should also consider the occurrence of unjustified risky trajectories either between close obstacles or around them with little space left. 
    
\noindent    An \textit{Unsafe Label} is assigned when UAV's minimum distance to obstacles falls below \textit{1m}, or it is in the range  $[1:3]m$ with a flight path relative to the obstacles that indicate a high collision risk.

\noindent    A \textit{Safe Label} is assigned in the other cases, i.e.,  when the UAV consistently maintains a safe distance from obstacles throughout its flight path, with no apparent risk.

\paragraph{Certainty Assessment.}
    Following the definition and description provided in Section \ref{sec:background-uncertainty} for \textit{decision uncertainty},
    a \textit{Certain Label} is assigned if the UAV exhibits a clear, consistent trajectory, indicating direct and smooth navigation.

    \noindent An \textit{Uncertain Label} is used when the UAV shows erratic or inconsistent movements in a certain segment of the flight path, suggesting indecisive or unpredictable navigation. Individual or infrequent changes due to initial unawareness of the drone about the surrounding environment are discarded, as they are indicative of an explorative, rather than uncertain, behavior.

\paragraph{Labeling Process.}
To ensure the robustness and objectivity of the dataset labels, three of the authors acted as validators and independently labeled the datasets.
Validators were provided with sample plots (similar to Figure \ref{fig:safety-matrix}) as visual examples of the four possible cases. Labeling guidelines were used to reinforce the standardization of the labeling process.
Each author, well-versed in UAV operational dynamics and the specific objectives of this study, independently reviewed the same set of flight data and assigned Safety and Certainty labels to the flight scenarios.
After individual labeling, the final label for each flight scenario was determined through majority voting among three validators to mitigate bias and ensure a balanced representation. To quantify the consistency and reliability of our manual labeling process, we compute and report the agreement rate among the validators. 

\subsubsection{Correlation Analysis}
Analyzing labeled flight datasets, we explore the correlation between unsafe and uncertain UAV behavior. Our research hypothesis justifies our analysis:
\begin{tcolorbox}[colback=gray!15!white,colframe=black, boxsep=2pt,left=3pt,right=3pt,top=3pt,bottom=3pt]
\textbf{Hypothesis}: Uncertain decisions by the UAV likely lead to unsafe states.
\end{tcolorbox}

\paragraph{Confusion Matrix (Safety vs. Certainty).}
We generate a confusion matrix aligning flight safety (safe or unsafe) with UAV decision certainty (certain or uncertain) to visually represent the distribution and overlap of these critical behaviors. We then compute \textit{agreement accuracy} from the confusion matrix, measuring the proportion of total observations where safety and certainty behaviors co-occur (or align): the UAV is both safe and certain or both unsafe and uncertain.
More formally,  we calculate two key conditional probabilities: \( p(\text{unsafe} | \text{uncertain}) \) and \( p(\text{uncertain} | \text{unsafe}) \). 
These probabilities express the chance of a flight being unsafe given uncertainty and vice versa, offering deeper insights into their relationship.

To enhance the statistical reliability of our results, we compute  \textit{Wilson's Confidence Interval} for proportions~\cite{Wilson1927} on the mentioned probabilities. This approach provides the confidence interval for the two probabilities we are interested in, indicating the range in which the true probability values fall with some confidence $\gamma$ (we adopt the commonly used value of $\gamma = 0.95$).

\subsection{RQ$_2$ \& RQ$_3$ - Uncertainty Detection and Unsafety Prediction}
\label{sec:methodology-rq2}
In RQ$_2$ and RQ$_3$, we leverage our hypothesis (from RQ$_1$) that detecting uncertainty at runtime serves as a proxy for predicting unsafe behavior. Using an anomaly detection approach, we identify decision uncertainty, used as a proxy for unsafety. We explain the preprocessing steps of flight data and how we employ autoencoders for automating the detection of UAV uncertainty. Lastly, we discuss the evaluation steps and metrics.

\subsubsection{Data Preprocessing}
\label{sec:preprocess}
\paragraph{Log Messages.}
The flight logs, detailed in section \ref{sec:backgd-logs}, contain extensive information requiring proper preprocessing. As we adopt a \textit{black-box} uncertainty detection approach, we focus solely on the logged output of the PX4-Avoidance module (see Figure \ref{fig:px4}), specifically, the \textit{Safe Waypoints} $\langle ts, x_s, y_s, z_s, r_s\rangle$, which represent high-level control signals steering the drone's flight trajectory. 

Our preliminary experiments using different subsets of these features ($\langle x_s$, $y_s$, $z_s, r_s\rangle$, $\langle x_s, y_s, z_s\rangle$ and $\langle r_s\rangle$) revealed that the \textit{"heading"  angle of the drone} ($r_s$) is the most reliable and significant source of information about its decision certainty. 
This can be explained by the fact that the heading angle strictly represents the viewpoint of our case study drone with a single forward-oriented camera. 
Changes in heading are vital for surveying in uncertain scenarios. We observed increased rotational movements in such situations, establishing a distinct pattern in the heading data as a strong indicator of anomalies. 
Moreover, adding position coordinates to the inputs made the trained model overfit the training data, even after proper preprocessing steps. 
So, in the context of our study, we preprocess the \textit{Safe Waypoints} data by considering only the heading angle $r$, excluding $x$, $y$, and $z$. 
By concentrating on this single black-box variable, we simplify our anomaly detection model, using a minimal set of values in a time window to capture key features indicative of uncertainty.

We implement a further preprocessing step on the resulting (<ts, r$_s$>) time series. 
PX4 works with angles in the range of [$-180^{\circ}$, $180^{\circ}$] (originally [$-\pi$, $\pi$] in radians). This results in huge jumps from $-180^{\circ}$ to $180^{\circ}$ ($-\pi$ to $\pi$) when the drone is constantly rotating. For instance, if a drone with an angle of 175$^{\circ}$ rotates 10$^{\circ}$ anti-clockwise, the resulting angle would be -175$^{\circ}$, with a difference between the two consecutive angles equal to 350$^{\circ}$ instead of 10$^{\circ}$. We apply a continuous transformation of the heading angle in the time series (if the difference between two consecutive values is bigger than $180^{\circ}$ ($\pi$), we deduct or add $360^{\circ}$ ($2\pi$) to the later one accordingly) to ensure that small rotations are always associated with small differences between consecutive angular measurements.

\paragraph{Windowed Dataset.}
To monitor drone behavior in real time, we segment time series data into windows. Each timestamp accesses only the current window data (and possibly previous ones), excluding future data, aligning with the runtime scenario. We create a windowed dataset from flight log messages, specifying window length and half-length overlap. The dataset includes columns for each time window:
\begin{itemize}
    \item \textit{start} and \textit{end} timestamps, and the window \textit{index};
    \item \textit{Win\_dist}: minimum distance of the drone to the obstacles during the window;
    \item Zero-centered array of heading angles $\langle \hat{r_1}, \ldots, \hat{r_n}\rangle$, obtained from the safe trajectory waypoint values $\langle r_s,x_s,y_s,z_s\rangle$ by subtracting the average value of $r$ within the window from each of the values (this ensures that different time series obtained when the drone is heading toward different directions are comparable with each other);
    \item \textit{Min\_dist}: minimum distance of the drone to the obstacles during the whole flight;
    \item \textit{Safety} and \textit{Certainty} states (from the manual labeling) of the drone.
\end{itemize}


    


Table \ref{tab:datasets} summarises the datasets we used in our experiments. \textit{train\_ds} and \textit{test1\_ds} are both obtained from the same test generation process using the extended version of \textit{Surrealist} we created. We split the test generation output into training and test sets with an 80-20\% ratio, the former used to train the anomaly detectors based on autoencoders, and the latter used in our experiments to empirically validate the approach. Specifically, we take test cases from the initial (first 9 iterations) and final (last 12 iterations) steps of the test generation process for the validation set to ensure the minimum similarity between the two sets.
To evaluate our approach, we use a dataset from the original \textit{Surrealist}'s study \cite{surrealist} referred to as \textit{test2\_ds} to ensure the approach works for flights generated with a different test generation approach.
For each dataset, we generated time windowed datasets using a \textit{window length of 5 seconds} and an \textit{overlap of 2.5 seconds} (decided empirically after initial experiments with window length from 2.5 up to 20 seconds).

    \begin{table}
    \caption{Experimental Datasets}
    \label{tab:datasets}
    \begin{minipage}{\columnwidth}
    \begin{center}
    \begin{tabular}{lcccc}
    \toprule
    \textbf{Dataset}   & \textbf{\#Tests$\times$Exec.} & \textbf{\#Flights} & \textbf{\#<1.5m}  &\textbf{\#Windows}  \\
    \midrule
    \textit{train\_ds}  & 86$\times$9   & 767   & 325 (42\%) & 37'810 \\ 
    \textit{test1\_ds}    & 21$\times$9   & 189   & 42 (22\%)  & 9'193 \\
    \textit{test2\_ds}   & 53$\times$10  & 542   & 199 (37\%) & 30'251 \\
    \bottomrule
    \end{tabular}
    \end{center}
    \end{minipage}
    \end{table}

\subsubsection{Runtime Uncertainty Detection}

\label{sec:uncertainty}
\begin{figure}
\centering
\includegraphics[width=\columnwidth]{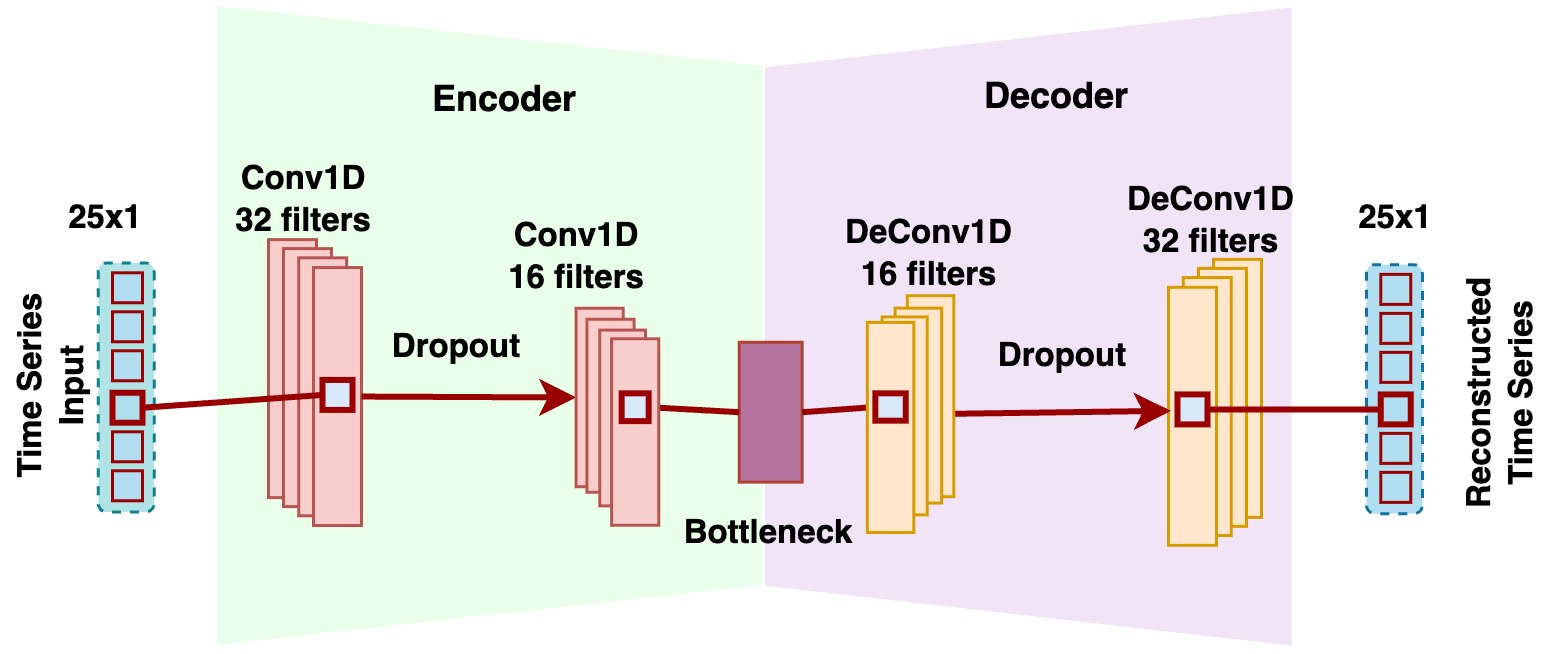}
\vspace{-1mm}
\caption{\textsc{Superialist}'s Autoencoder Architecture}
\label{fig:autoencoder}
\vspace{-5mm}
\end{figure}

We developed \textsc{Superialist}, a real-time uncertainty detection approach for UAVs, using an autoencoder\footnote{The main alternative to auto-encoders for anomaly detection is the use of clustering and the computation of the distance between a new trajectory and the cluster centroids, where a high distance would indicate a potential anomaly. However, clustering approaches to anomaly detection rely on the definition of a proper distance metric that is capable of distinguishing uncertain from certain behaviors. On the contrary, autoencoders do not require any distance metric. Moreover, they can extract latent features of time series in a non linear way. This is not possible with clustering.} to learn high-level control system behavior in an offline phase, enabling real-time detection of anomalies, i.e., uncertainty instances during runtime.

\paragraph{Autoencoder.}
An Autoencoder, a type of artificial neural network, is trained through unsupervised learning. It learns to produce outputs that accurately reconstruct its original inputs (effectively implementing the identity function). This process involves two main components (demonstrated in Figure \ref{fig:autoencoder}): an encoder and a decoder. The encoder compresses input data into a lower-dimensional representation known as the ``latent space'', while the decoder reconstructs the data to its original dimensionality \cite{thill2021temporal}. Autoencoders aim to learn a representation (encoding) for a set of data, often for tasks like dimensionality reduction or anomaly detection.

To demonstrate the feasibility of using autoencoders for uncertainty detection in our context, we employ a convolutional autoencoder with a simple, common architecture for anomaly detection in time-series data~\cite{autoencoder-tutorial}, depicted in Figure \ref{fig:autoencoder}. The input comprises a time window of heading values, encompassing 25 data points across 5 seconds. The encoder consists of two 1D convolutional layers (with a kernel size of 3) for feature extraction, interspersed with dropout layers to prevent overfitting. Subsequently, the decoder employs two 1D convolutional transpose (DeConv) layers to gradually reconstruct the signal to its original dimensionality. The model, compiled with the Adam optimizer and utilizing mean squared error for loss, effectively learns to detect anomalies by identifying significant deviations in the reconstruction of UAV control signals from their original patterns (i.e., the reconstruction error tends to be higher for anomalous than for nominal data).

\paragraph{Training.}
Our training dataset includes normal, safe, and certain flights as well as anomalous, uncertain, unsafe, and crashing flights. In anomaly detection applications, it is generally recommended to, when feasible, train the auto-encoders exclusively with `nominal' data, to make sure the trained model does learn to reconstruct them, while producing a high reconstruction error on anomalies~\cite{thill2021temporal}. 
We empirically realized that the drone keeps a safe distance of over 3.5m to the obstacles when not challenged with complicated obstacle placements.
So, to automatically distinguish the nominal data representing normal and safe operational states of UAVs, we filter the time windows in unlabeled \textit{train\_ds} where the distance from obstacles exceeds 3m in the next 50 seconds.  This ensures that we exclude potential anomalous states close to the obstacles while keeping most of the available nominal states in the training dataset.
This training approach allows the autoencoder to learn typical patterns and characteristics of normal UAV behavior. The model optimizes its weights and biases during this phase to minimize the difference between input and reconstruction, measured by a metric known as reconstruction loss.

\paragraph{Inference.}
After the training phase, the autoencoder is utilized to analyze real-time data generated by the UAV. The autoencoder maintains a low reconstruction loss when the UAV operates under standard and expected conditions. This low loss indicates a high fidelity between the input data and its reconstructed output. However, the reconstructed output significantly diverges from the input when the UAV faces atypical scenarios, malfunctions, or conditions not encountered during training. This increased divergence leads to a higher reconstruction loss, serving as a flag for potential anomalies. We leverage this inherent property of autoencoders to detect anomalies in the control signals of UAVs. By continuously monitoring the reconstruction loss of control signal data during the flight, we can pinpoint instances where the UAV's behavior deviates from the 'normal' patterns learned during training.

\subsubsection{Evaluation}
Following the training phase, the model is rigorously assessed using two separate, labeled datasets. These datasets are specifically chosen to evaluate the model's efficacy in identifying uncertain behaviors in diverse flight scenarios using the following metrics:

\noindent    \textit{- Confusion Matrix:} Visually represents the model's predictions against the ground truth certainty labels from manual labeling. The confusion matrix is a $2\times 2$ matrix consisting of the following rows: $[TP, FP]$ (true positives/false positives, i.e., correctly/incorrectly predicted uncertain states) and $[FN, TN]$ (false negatives/true negative, i.e., incorrectly/correctly predicted certain states). 

\noindent    \textit{- Precision:} Measures the proportion of actual uncertain cases among all cases labeled as uncertain by the model ($p = TP / (FP + TP) $).

\noindent    \textit{- Recall:} Reflects the proportion of actual uncertain cases that were correctly identified by the model among those to be recovered ($r = TP / (FN + TP)$).

\noindent    \textit{- F1 Score:} Provides a harmonic mean of precision and recall, offering a balance between the two metrics ($F_1 = 2pr/(p+r)$).

We report the same metrics considering \textit{Safety} as the ground truth, i.e., evaluating our approach for predicting future safety violations instead of detecting uncertainty to answer RQ$_3$.  


\section{Experimental Results}
\label{sec:results}

\subsection{RQ$_1$ - Unsafety vs. Uncertainty}

\begin{table}[]
\centering
\caption{Unsafety - Uncertainty Confusion Matrix}
\label{tab:rq1-conf}
\begin{minipage}{\linewidth}
\centering
\caption*{\textit{test1\_ds}}
\vspace{-3mm}
\begin{tabular}{|c|c|c|c|}
\hline
\textbf{Unsafe} & \multicolumn{2}{c|}{\textbf{Uncertain}} & Sum \\ \cline{2-3} 
\multicolumn{1}{|c|}{} & True & False & \multicolumn{1}{c|}{} \\  \hline 
True   & 16 (8.5\%)     & 9 (4.8\%)     & 25 (13.2\%) \\ \hline 
False  & 16 (8.5\%)     & 148 (78.3\%)  & 164 (86.8\%) \\ \hline 
Sum    & 32 (16.9\%)    & 157 (83.1\%)  & 189 (100\%) \\ \hline
\end{tabular}
\end{minipage}
\begin{minipage}{\linewidth}
\centering
\vspace{+2mm}
\caption*{\textit{test2\_ds}}
\vspace{-3mm}
\begin{tabular}{|c|c|c|c|}
\hline
\textbf{Unsafe} & \multicolumn{2}{c|}{\textbf{Uncertain}} & Sum \\ \cline{2-3} 
\multicolumn{1}{|c|}{} & True & False & \multicolumn{1}{c|}{}\\ \hline
True   & 123 (22.7\%)   & 15 (2.8\%)    & 138 (25.5\%) \\ \hline
False  & 43 (7.9\%)     & 361 (66.6\%)  & 404 (74.5\%) \\ \hline
Sum    & 166 (30.6\%)   & 376 (69.4\%)  & 542 (100\%) \\ \hline
\end{tabular}
\end{minipage}
\end{table}

\begin{table}[]
\centering
\caption{Statistical Metrics for Safety - Certainty Relation}
\label{tab:rq1-metrics}
\begin{tabular}{lcc}
\toprule
\textbf{Metric} & \textbf{\textit{test1\_ds}} & \textbf{\textit{test2\_ds}} \\
\midrule
Agreement Accuracy (\%) & 86.8\% & 89.3\% \\
$p(\text{unsafe} \mid \text{uncertain})$ & 50\% & 74.1\% \\
Conf interv (unsafe $\mid$ uncertain) & [33.6, 66.4]\% & [66.9, 80.2]\% \\
$p(\text{uncertain} \mid \text{unsafe})$ & 64\% & 89.1\% \\
Conf interv (uncertain $\mid$ unsafe) & [44.5, 79.8]\% & [82.8, 93.3]\% \\
\bottomrule
\end{tabular}
\end{table}

We manually labeled \textit{test1\_ds} and \textit{test2\_ds}. \textit{train\_ds} was not labeled since it was used to train the prediction model in an unsupervised way. While most of the labeling disagreements among the three validators were about the flights at the borderline of safe/unsafe or certain/uncertain, the disagreement rate was higher for \textit{test1\_ds} (10.5\%) compared to \textit{test2\_ds} (6.5\%), suggesting the existence of more borderline cases.

The confusion matrix in Table \ref{tab:rq1-conf} illustrates the relationship between safety and uncertainty within our manually labeled datasets. Specifically, \textit{test1\_ds} comprises 13.2\% unsafe flights, whereas \textit{test2\_ds} contains approximately twice that percentage at 25.5\%. 
\textit{test2\_ds} has also a higher proportion of uncertain decisions at 30.6\%, in comparison to \textit{test1\_ds} at 16.9\%. 
This discrepancy arises from our customization of the \textit{Surrealist} implementation: while the original implementation results in \textit{test2\_ds}, our modified version leading to \textit{test1\_ds} aims to foster increased diversity, with safety violation not being its sole objective.
Interestingly, our manual safety labels are only partially aligned with a simple label based on the 1.5m safety distance threshold suggested by the previous work~\cite{surrealist}. They disagree in 9\% of the flights in \textit{test1\_ds} and 12\% in \textit{test2\_ds}, mainly due to the cases with a distance lower than 1.5m that were manually labeled as safe.

\begin{figure*}
\centering
\vspace{-3mm}
\includegraphics[width=\textwidth]{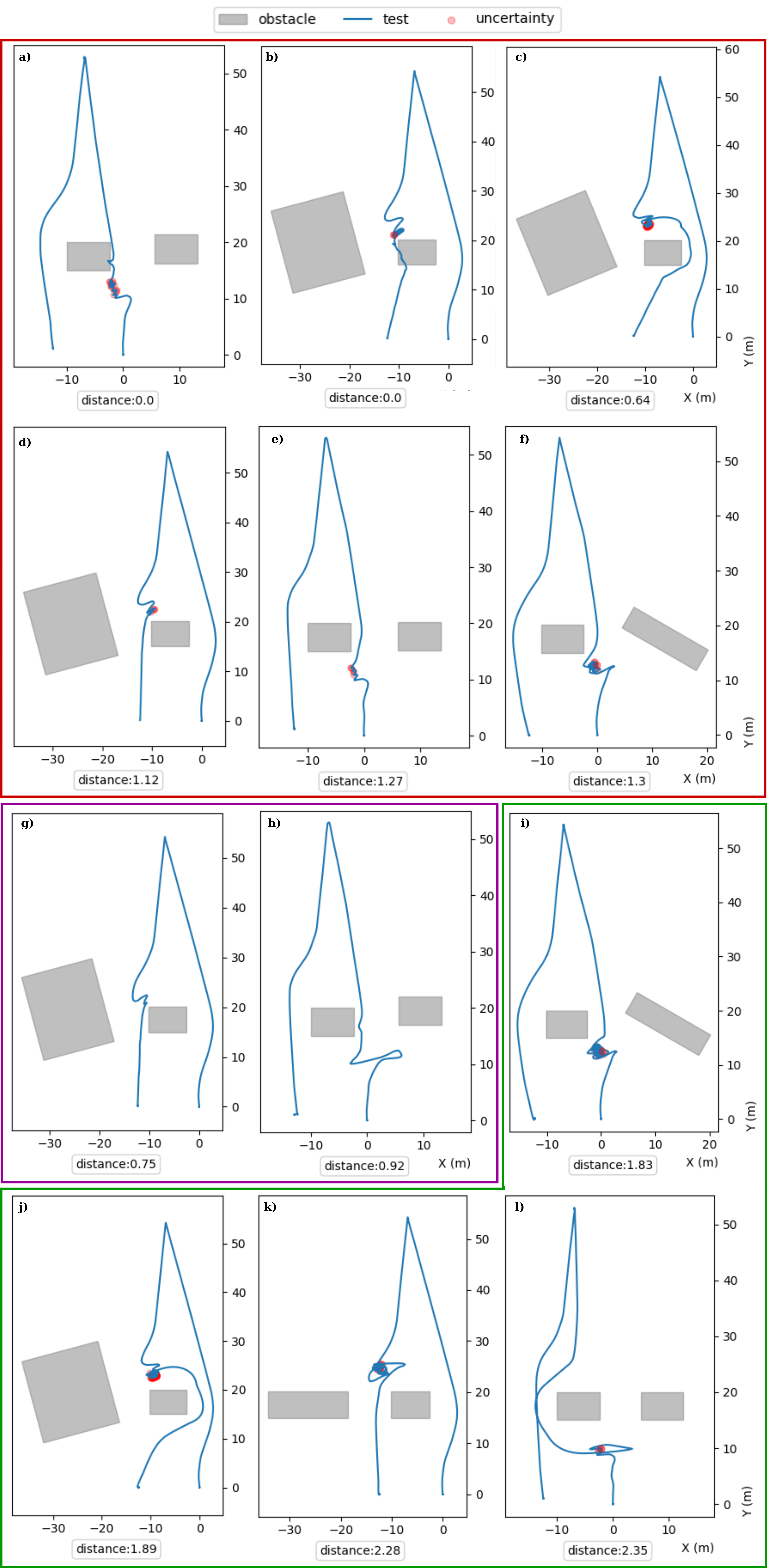}
\vspace{-5mm}
\caption{Uncertainty Detection Examples ([a-f]: unsafe, uncertain $\mid$ [g,h]: unsafe, certain $\mid$ [i-l]: safe, uncertain)}
\label{fig:detection-examples}
\end{figure*}

As shown in Table \ref{tab:rq1-metrics}, a substantial level of agreement is evident between the two labels for both datasets, with approximately 87\% to 89\% of flights falling into either the categories of both being unsafe and uncertain or both being certain and safe.
Figure \ref{fig:detection-examples}[a-f] illustrates examples of unsafe flights, including crash scenarios [a, b], exhibiting varying degrees of uncertainty preceding the unsafe states.
Interestingly, in all evaluated scenarios, the first signs of uncertainty are always observed at least a few seconds before the unsafe states (detailed in Section \ref{sec:results-rq3}).
An illustration of safe and certain scenarios is already presented in Figure \ref{fig:safety-matrix}[a].

Based on $p(\text{unsafe}\mid\text{uncertain})$ (see Table \ref{tab:rq1-metrics}), we can notice that
between 50\% to 74\% of the uncertain decisions are leading to unsafe states (Figure \ref{fig:detection-examples}[a-f]), which also indicates that 26\%  to 50\% of the uncertain decisions are not leading to unsafe states. Figure \ref{fig:detection-examples}[i-l] shows four examples where the drone is uncertain in path planning decisions for a few seconds, but it can eventually find a safe path.
    \begin{tcolorbox}[colback=gray!15!white,colframe=black, boxsep=2pt,left=3pt,right=3pt,top=3pt,bottom=3pt]
    \textbf{Finding 1}: 
    Up to 74\% of the uncertain UAV decisions lead to unsafe flight states shortly, which leaves at least 26\% of the uncertain decisions not correlated with unsafety. 
    \end{tcolorbox}
Based on $p(\text{uncertain}\mid\text{unsafe})$ (see Table \ref{tab:rq1-metrics}), our analysis reveals that 64\% to 89\% of unsafe states also manifest some signs of uncertainty (refer to Figure \ref{fig:detection-examples}[a-f] for some examples). This suggests that a notable portion, ranging from 11\% to 36\%, of unsafe states occur in flights where no significant uncertainty is evident. In Figure \ref{fig:detection-examples}[g,h], we present two sample flights where predominantly certain decisions lead to unsafe flight states, i.e., the drone experiences failures while operating with high certainty. This subset of unsafe scenarios may not be effectively predicted using an uncertainty-based approach.

    \begin{tcolorbox}[colback=gray!15!white,colframe=black, boxsep=2pt,left=3pt,right=3pt,top=3pt,bottom=3pt]
    \textbf{Finding 2}: 
    Up to 89\% of unsafe states exhibit significant signs of uncertainty before the incident, while over 11\% of unsafe states occur in flights where uncertainty is not observed.
    \end{tcolorbox}

The Wilson confidence intervals reported for each of the above conditional probabilities consist of relatively bigger intervals for \textit{test1\_ds}, which is mostly due to the smaller population size of this dataset. Overall, the ranges of the 95\% confidence intervals indicate that, despite the error affecting our probability estimates, our conclusions (in particular, Findings 1 and 2) remain valid and reliable.


\subsection{RQ$_2$ - Uncertainty Detection}


Our proposed approach is based on anomaly detection using an autoencoder. Following the steps in section \ref{sec:methodology-rq2}, we trained the model with 17,191 nominal data rows (time windows) filtered from \textit{train\_ds}.
The training phase reached convergence after 246 epochs. 

\paragraph{Threshold.}
\begin{figure}
\centering
\includegraphics[width=\columnwidth]{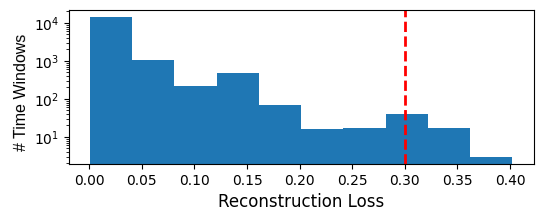}
\caption{Histogram of Nominal Data Reconstruction Loss}
\label{fig:loss-hist}
\end{figure}
Establishing an optimal threshold for the reconstruction loss is a crucial element of our methodology. This threshold serves to distinguish between normal and anomalous UAV behavior, i.e., certain and uncertain decisions. 
We empirically realized that during the flights, a sudden peak in reconstruction loss is expected and normal when initiating common maneuvers such as intentional turns at mission waypoints or switching flight modes. To accommodate this, instead of considering the reconstruction loss in single time windows, we enhance our methodology by computing the \textit{mean reconstruction loss} over $n = 4$ consecutive time windows. This modification provides a more precise representation of the UAV's operational condition, mitigating minor, temporary fluctuations of the reconstruction loss and reducing the false positives.

Since there is no established standard or guideline on the \textit{accepted} or normal level of decision uncertainty for UAVs, we need to empirically determine a threshold based on our datasets. Hence, we rely on the histogram of the reconstruction loss derived from the nominal training data. Figure \ref{fig:loss-hist} visually represents the frequency distribution of different loss levels observed during nominal UAV operations.
For the majority of time windows, the reconstruction loss is below 0.05, as depicted in the figure (note the logarithmic scale). However, there is a significant drop in the count for values exceeding 0.2, with the maximum value reaching 0.4. Given the expectation of a small number of anomalies in the training data that our filter may not have identified, we have decided to set the anomaly detection (i.e., uncertainty detection) threshold at $\theta = 0.3$. Moreover, given the importance and criticality of early detection of uncertain decisions for preventing unsafe and crashing scenarios, we are interested in increasing the True Positive Rate (i.e., correctly detected uncertainties) of the approach while accepting a limited possibility of increasing its False Positive Rate (i.e., false alarms) as a consequence.

This approach ensures that the threshold is grounded in the actual data distribution, allowing us to effectively identify and capture anomalous behaviors that deviate significantly from the norm. Using this threshold, only 39 time windows of the nominal dataset, belonging to 4 different flight logs, were labeled as uncertain. Manual analysis revealed that all of them are indeed uncertain flights that were not filtered from the training data, hence giving a false positive rate of 0\% on the training data.

\begin{figure}
\centering
\vspace{-3mm}
\includegraphics[width=0.95\columnwidth]{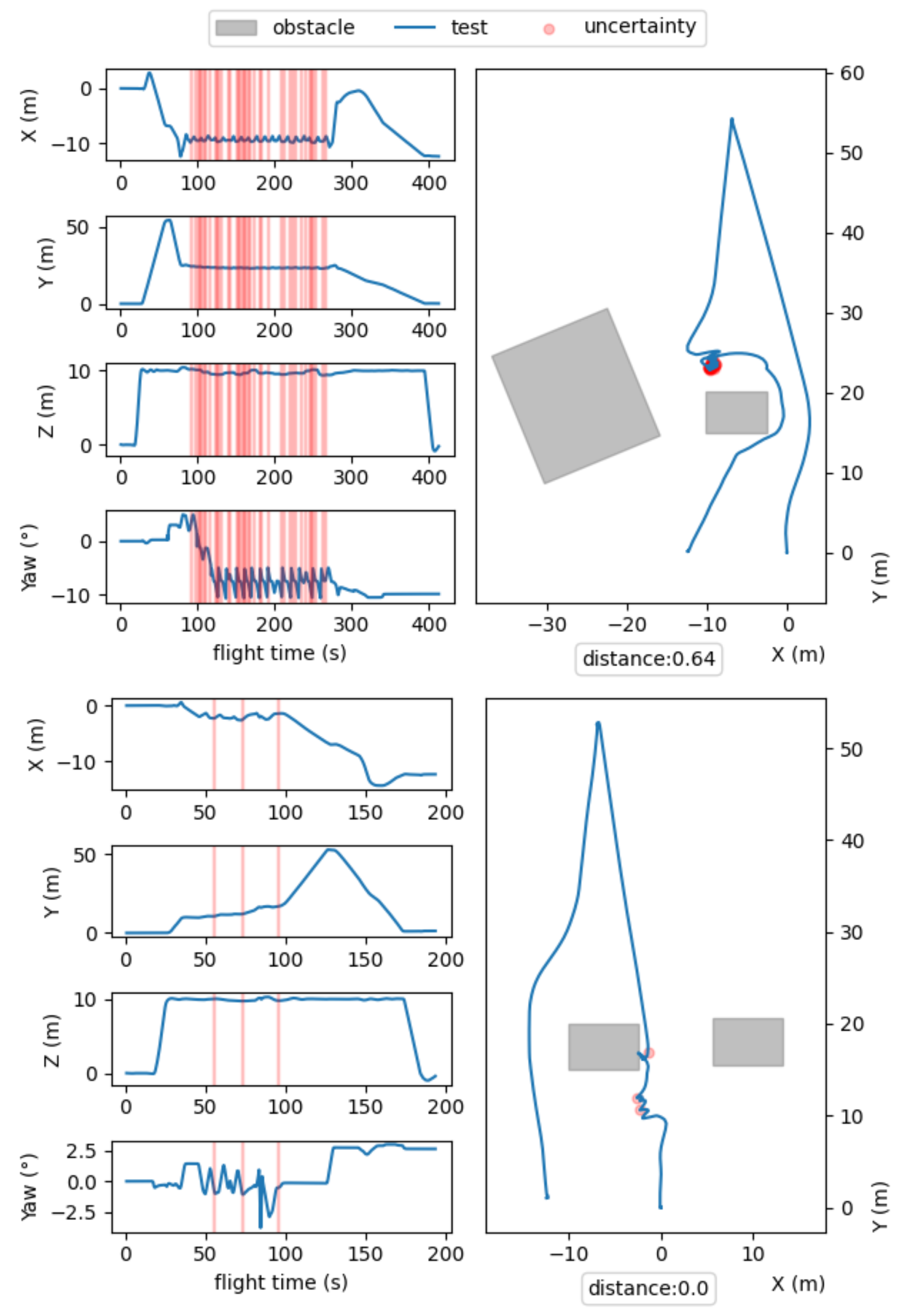}
\vspace{-2mm}
\caption{Detected Uncertainty Over Flight Time}
\label{fig:sample-flags}
\vspace{-5mm}
\end{figure}

Figure \ref{fig:sample-flags} illustrates how \textsc{Superialist} successfully learned to identify anomalous patterns in the control signals, particularly in the heading angle of the drone (yaw). These patterns correspond to observable uncertainty and instability in the drone's behavior.
These uncertainty patterns are early indicators of potentially unsafe states ahead (i.e., before the crash). In the specific flight analyzed at the top, the initial uncertainty alarms (highlighted with red lines in time-wise plots on the left and with red spots in the X-Y plot on the right) are triggered approximately 90 seconds after takeoff. Subsequently, the drone enters an unsafe area around the 300-second mark. Throughout this period, the drone hovers within a confined space, continuously changing yaw (heading angle) as it searches for a suitable path forward.
In light of our warning mechanism, a precautionary corrective process can be initiated to avert the drone from entering the unsafe area. Alternatively, a human pilot could be notified to intervene and safely navigate the drone in this critical state, transitioning control back to autopilot when conditions are deemed safe.
This is also true for the second example at the bottom, where a substantially less evident uncertainty pattern, first detected at around the 50-second mark, leads to an eventual crash to the obstacle about 90 seconds after taking off.

\begin{table}[]
\centering
\caption{Uncertainty Detection Confusion Matrix}
\label{tab:rq2-conf}
\begin{minipage}{.5\linewidth}
\centering
\caption*{test1\_ds}
\vspace{-3mm}
\begin{tabular}{|c|c|c|}
\hline
\textbf{Predicted} & \multicolumn{2}{c|}{\textbf{Uncertain}} \\ \cline{2-3} 
\multicolumn{1}{|c|}{} & True & False \\ \hline
True    & 27    & 3 \\ \hline
False   & 5     & 154 \\ \hline
\end{tabular}
\end{minipage}%
\begin{minipage}{.5\linewidth}
\centering
\caption*{test2\_ds}
\vspace{-3mm}
\begin{tabular}{|c|c|c|}
\hline
\textbf{Predicted} & \multicolumn{2}{c|}{\textbf{Uncertain}} \\ \cline{2-3} 
\multicolumn{1}{|c|}{} & True & False \\ \hline
True    & 155   & 7 \\ \hline
False   & 11    & 369 \\ \hline
\end{tabular}
\end{minipage}
\end{table}

\begin{table}[]
\centering
\caption{Model Evaluation Metrics by Ground Truth Label and Dataset}
\label{tab:rq2-metrics}
\begin{tabular}{llcccc}
\toprule
\textbf{Label} & \textbf{Dataset}       &  \textbf{Acc.} & \textbf{Prec.} & \textbf{Recall} & \textbf{F1} \\
\midrule
\multirow{2}{*}{\textbf{Uncertainty}}   & \textit{test1\_ds}  & 95.8\%& 90\%  & 84.4\%  & 87.1\% \\
                                        & \textit{test2\_ds}  & 96.7\%& 95.7\%& 93.4\%& 94.5\% \\
\midrule    
\multirow{2}{*}{\textbf{Unsafety}}      & \textit{test1\_ds}  & 84.7\%& 43.3\%& 52\%  & 47.3\% \\
                                        & \textit{test2\_ds}  & 88.9\%& 74.1\%& 87\%  & 80\% \\
\bottomrule
\end{tabular}
\end{table}

Table \ref{tab:rq2-conf} presents the confusion matrix for uncertainty detection using \textsc{Superialist}, while aggregated performance metrics are detailed in Table \ref{tab:rq2-metrics}. Values indicate consistently high overall accuracy and precision (90-96\%), confirming the reliability of \textsc{Superialist} in detecting anomalies.
The substantial recall rate of 93\% for \textit{test2\_ds} underscores the tool's effectiveness in identifying uncertain decisions. However, a slightly lower recall of 84\% for \textit{test1\_ds} is noted, potentially attributed to the dataset's greater scenario diversity.
These metrics collectively show that \textsc{Superialist} properly identifies the patterns that characterize uncertainty in drone path planning decisions.
\begin{tcolorbox}[colback=gray!15!white,colframe=black, boxsep=2pt,left=3pt,right=3pt,top=3pt,bottom=3pt]
\textbf{Finding 3}: 
\textsc{Superialist} effectively detects anomalies in the high-level control signals of UAVs, enabling cost-effective real-time identification of decision uncertainties. With a precision of up to 95\% and a recall of 93\%, autoencoders stand as a reliable black-box solution for monitoring UAV decision-making.
\end{tcolorbox}
\begin{table}[]
\centering
\caption{Unsafety Prediction Confusion Matrix}
\label{tab:rq3-conf}
\begin{minipage}{.5\linewidth}
\centering
\caption*{test1\_ds}
\begin{tabular}{|c|c|c|}
\hline
\textbf{Predicted} & \multicolumn{2}{c|}{\textbf{Unsafe}} \\ \cline{2-3} 
\multicolumn{1}{|c|}{} & True & False \\ \hline
True    & 13     & 17 \\ \hline
False   & 12     & 147 \\ \hline
\end{tabular}
\end{minipage}%
\begin{minipage}{.5\linewidth}
\centering
\caption*{test2\_ds}
\begin{tabular}{|c|c|c|}
\hline
\textbf{Predicted} & \multicolumn{2}{c|}{\textbf{Unsafe}} \\ \cline{2-3} 
\multicolumn{1}{|c|}{} & True & False \\ \hline
True    & 120   & 42 \\ \hline
False   & 18    & 362 \\ \hline
\end{tabular}
\end{minipage}
\end{table}

\subsection{RQ$_{3}$ - Unsafety Prediction}
\label{sec:results-rq3}

Figure \ref{fig:detection-examples} illustrates several instances of detected uncertainty during flights. The top row ([a-f]) plots depict flights manually labeled as both unsafe and uncertain. Our approach successfully identifies uncertainty in the highlighted areas (red spots), and notably, this detection precedes the actual safety violations, allowing for early prediction.
Indeed, our analysis reveals that, on average, the first warning is raised 50.2 seconds before the drone reaches <1m distance to the obstacles in \textit{test1\_ds}, and 42.4 seconds in \textit{test2\_ds}, when the drone was on average 3.45m and 3.72m away from the obstacles. Considering the negligible preprocessing and auto-encoder inference time, the drone pilot, or any automated self-healing approach, has a very large time and distance window to react (See examples in Figure \ref{fig:sample-flags}).

Plots [i-l] showcase flights labeled as uncertain but ultimately safe. In these instances, \textsc{Superialist} correctly identifies uncertainty in the UAV's decisions, yet the drone navigates safely in subsequent seconds. This phenomenon contributes to the observed lower precision (43-74\%) when predicting unsafety instead of uncertainty (refer to Tables \ref{tab:rq2-metrics} and \ref{tab:rq3-conf}).
This is consistent with our Findings 1 and 2: black-box uncertainty is not sufficient to predict unsafety in a non-negligible proportion of the cases. Hence, even a perfect uncertainty predictor would have suboptimal performance if employed to predict unsafety.
Indeed, $p(\text{unsafe}\mid\text{uncertain})$ (Table \ref{tab:rq1-metrics}) sets the precision upper bound to 50\% and 74\% in the two datasets.
 
Plots [g, h] are labeled as unsafe but certain. Unlike the previous examples, although the drone adjusts its flight direction multiple times to optimize its path, it eventually selects the final route without hesitation. Such cases are aligned with our previous Findings 1 and 2 (unsafety correlates mildly with uncertainty). This contributes to the lower recall observed for \textit{Unsafety} labels (52-87\%) 
with the upper bound being 64\% and 89\% in the respective datasets considering $p(\text{uncertain}\mid\text{unsafe})$ in Table \ref{tab:rq1-metrics}.

\begin{tcolorbox}[colback=gray!15!white,colframe=black, boxsep=2pt,left=3pt,right=3pt,top=3pt,bottom=3pt]
\textbf{Finding 4}: 
Anomalies in control signals are moderately effective for predicting unsafe drone states, with up to 74\% precision and 87\% recall. However, given our previous findings on the correlation between uncertainty and unsafety, we know that even a perfect uncertainty detector would perform suboptimally for unsafety state prediction. More factors, including white-box ones,  should probably be considered for a (less cost-effective but) more precise and reliable safety supervisor.
\end{tcolorbox}

\subsection{Implications and Lessons Learned}\label{sec:lessons-learnt}

\begin{figure}
\centering
\includegraphics[width=0.95\columnwidth]{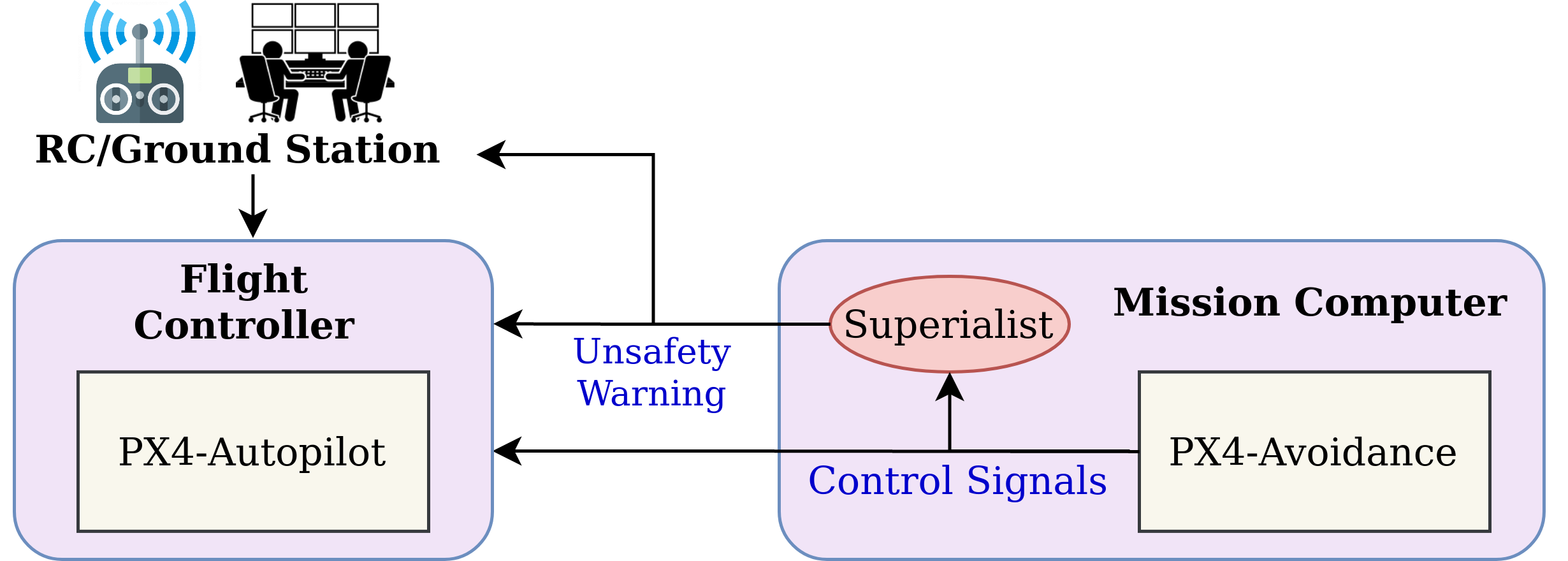}
\caption{Suggested Deployment Architecture for \textsc{Superialist}} 
\label{fig:superialist}
\vspace{-5mm}
\end{figure}

The relationship between uncertainty and safety in UAVs is multifaceted. Safety, influenced by various factors including proximity to the surrounding obstacles and environmental conditions, does not always correlate with UAV uncertainty states. Thus, our proposed uncertainty estimation approach may not always directly address safety in UAV test scenarios.
Nevertheless, an uncertainty-based monitoring system with a wide reaction window can still play a crucial role in enhancing the overall system safety by triggering autonomous self-healing mechanisms to recover the drone or alerting human pilots to potential risks, even if intervention is not always necessary.  
This is even more crucial considering the current regulations in commercial deployment of autonomous UAVs (e.g., for delivery tasks) that mandates the manual monitoring of the flights by remote pilots, where each pilot may monitor up to 20 concurrent flights~\cite{matternet}.
Figure \ref{fig:superialist} demonstrates our suggested deployment architecture to use \textsc{Superialist} (after training the anomaly detection model) on board of PX4-powered UAVs, enabling sending real-time safety warnings to the autopilot system and the remote pilots. This signal can also be integrated in a dynamic safety assurance framework~\cite{ryan2024dynamic} and trigger configuration changes to maintain safety of the system. 

To improve a black-box uncertainty detection approach for predicting unsafety, a trade-off is necessary—sacrificing "efficiency" and "generalizability" in favor of "effectiveness". 
While our lightweight (i.e., requiring limited processing power on board of UAVs), \textit{black-box} approach relying only on \textit{minimal trajectory planning data} (outputs of the path planning algorithm) can be easily deployed for alternative obstacle avoidance algorithms, as well as other UAV and autonomous robotic systems, any \textit{white-box} alternative will be inherently bound to a specific use case.

Future research should explore non-trivial gray-box approaches to enhance unsafety prediction in real-time scenarios while considering the processing limitations.
A promising research direction is combining our black-box approach with other white-box approaches, leveraging internal understandings of the path planning algorithm in addition to real-time data samples from cameras and various sensors.

\subsection{Threats to Validity}

\subsubsection{Internal Validity}
    Using a Convolutional Autoencoder to detect uncertain behaviors in UAVs might introduce model-specific biases. 
    If the training data is not representative of real-world scenarios, the model's generalization ability may be limited.
    To limit this threat, we rely on a dataset of diverse flight scenarios in the training data, while the evaluation data includes data from more realistic flights from previous work~\cite{surrealist}. 
    Also, manually labeling flight scenarios as safe/unsafe and certain/uncertain is subject to human error and subjective judgment. Despite providing guidelines to labelers to limit this threat, there might be inconsistencies in the interpretation and application of these guidelines. 
\subsubsection{External Validity}
    The datasets used for training and testing our model were derived from simulated UAV flights. The extent to which our findings can be generalized to real-world UAV operations is uncertain. The simulation may not fully capture variations in environmental conditions, UAV models, and operational scenarios in the real world. However, simulation-based testing is considered a best practice before moving to field testing.
    Moreover, the range of flight scenarios, environmental conditions, and obstacle configurations in our simulations may not represent all possible real-world challenges that UAVs might encounter. More specifically, for simplicity, we only considered a flight mission with three waypoints, two box-shaped obstacles, and without any wind. This limitation could affect the applicability of our results to more complex or unforeseen flight conditions, and to uncertain control decisions due to factors other than obstacles (e.g., wind). 


\subsubsection{Conclusion Validity}
    The statistical methods employed to analyze the data, including confusion matrices and conditional probabilities, need to be robust to ensure the validity of our conclusions. We used adequate sample sizes and statistical tests to avoid erroneous interpretations of the results. 
    In particular, we computed Wilson’s confidence interval for proportions when measuring \( p(\text{unsafe} | \text{uncertain}) \) and \( p(\text{uncertain} | \text{unsafe}) \), with confidence level set to 0.95. We also repeated each test 9 times in the simulator, to minimize the effect of non-determinism on our results.


\section{Conclusion and Future Work}
\label{sec:conclusion}
In this study, we explored the relationship between decision uncertainty and flight safety in UAV operations. Our empirical study focused on real-time detection of uncertain and unsafe behaviors, through the analysis of simulated UAV flight data. Key findings indicate a moderate correlation between uncertain decision-making and unsafe flight outcomes, with the Auto-encoder model demonstrating high precision and recall in uncertainty detection.  
Despite the lower precision in unsafety prediction, our safety monitoring approach with a wide reaction window can help to enhance safety in UAV operations by triggering fail-safe mechanisms, alerting human pilots, or integration into a dynamic safety assurance framework.

This research not only sheds light on the critical aspects of UAV safety but also opens ways for more advanced, data-driven safety measures in autonomous flight systems.
In future work, we aim to focus on identifying additional features, such as drone's speed and white/gray box indicators of uncertainty, to enhance unsafety prediction.
For instance, white box indicators of decision uncertainty could be obtained by inserting probes into the obstacle avoidance module. 
Moreover, we aim to extend and further evaluate our approach in dynamic environments (i.e., including moving obstacles), windy weather, in the field (i.e., real-world tests), and in other robotic domains.



\section*{Declarations}

\paragraph{Funding:} We thank the Horizon 2020 (EU Commission) support for the \href{https://www.innoguard.eu/index.html}{project InnoGuard}, 
Marie Sktodowska-Curie Actions Doctoral Networks (HORIZON-MSCA-2023-DN),  and the SNSF project entitled "SwarmOps: Human-sensing based MLOps for Collaborative Cyber-physical systems" (Project  No. 200021\_219732). We also thank the Hasler Foundation for the projects  "Bridging the Reality Gap in Testing Unmanned Aerial Vehicles" (Project No.23064) and "Advancing Unmanned Aerial Vehicles Reliability and Societal Trust Through Integrated Testing and Formal Verification" (Project No.2025-02-27-311). 

\paragraph{Data Availability:} All datasets, plots, and code needed to replicate all aspects of our work are included in the Replication Package \cite{RP} (\url{https://zenodo.org/doi/10.5281/zenodo.10393522}). 

\paragraph{Author Contributions:}
\begin{itemize}
\item\textit{Sajad Khatiri}: Conceptualization, Methodology, Software, Data Curation, Validation, Investigation, Writing – Original Draft
\item \textit{Fatemeh Mohammadi Amin}: Methodology, Software, Validation, Data Curation, Visualization, Writing – Original Draft
\item \textit{Sebastiano Panichella}: Conceptualization, Methodology, Data Curation, Writing – Review \& Editing, Supervision, Project Administration, Funding Acquisition
\item \textit{Paolo Tonella}: Conceptualization, Methodology,  Writing – Review \& Editing, Supervision
\end{itemize}

\paragraph{Conflict of Interest:} The authors declared that they have no conflict of interest.

\paragraph{Ethical Approval:} not applicable. 

\paragraph{Informed Consent:} not applicable. 

\paragraph{Clinical Trial Number:} not applicable.

\bibliographystyle{spmpsci}      
\bibliography{main}   

\end{document}